\documentclass[12pt,]{article}
\usepackage{authblk}
\usepackage{fullpage}
\usepackage{amssymb,amsmath}
\usepackage[utf8x]{inputenc}
\usepackage[T1]{fontenc}
\usepackage{siunitx}
\usepackage{caption}
\usepackage{subcaption}
\usepackage{rotating}
\usepackage[version=3]{mhchem}
\usepackage{authblk}

\usepackage[numbers,super]{natbib}

\usepackage{setspace}
\usepackage[unicode=true]{hyperref}
\hypersetup{breaklinks=true,
            bookmarks=true,
            colorlinks=true,
            linkcolor=blue,
            pdfborder={0 0 0}}
\usepackage{graphicx}
\graphicspath{{figures/}}
\makeatletter
\def\ScaleWidthIfNeeded{%
 \ifdim\Gin@nat@width>\linewidth
    \linewidth
  \else
    \Gin@nat@width
  \fi
}
\def\ScaleHeightIfNeeded{%
  \ifdim\Gin@nat@height>0.9\textheight
    0.9\textheight
  \else
    \Gin@nat@width
  \fi
}
\makeatother
\setkeys{Gin}{width=\ScaleWidthIfNeeded,height=\ScaleHeightIfNeeded,keepaspectratio}%

\title{Ultra-high-gain water-window X-ray laser driven by plasma photocathode wakefield acceleration}

\author{L. H. A. Berman$^{1,2*}$, D. Campbell$^{1,2}$, E. Hartmann$^{3}$, T. Heinemann$^{3}$,  T. Wilson$^{5}$, B. Hidding$^{3*,1}$ and A. F. Habib$^{1,2*}$}
\affil{$^1$ Department of Physics, Scottish Universities Physics Alliance, University of Strathclyde, Glasgow, UK.
\\$^2$ Cockcroft Institute, Sci-Tech Daresbury, UK. 
\\$^3$ Institute for Laser and Plasma Physics, Heinrich Heine University
D{\"u}sseldorf, D{\"u}sseldorf, Germany
\\$^5$ Theoretical Physics I, Heinrich Heine University
D{\"u}sseldorf, D{\"u}sseldorf, Germany
\\$^*$ Corresponding author(s): lily.berman@strath.ac.uk, bernhard.hidding@hhu.de and ahmad.habib@strath.ac.uk}

\date{\today}
\begin{document}
\maketitle
\begin{abstract}
X-ray free-electron lasers are large and  complex machines, limited by electron beam brightness. Here we show through start-to-end simulations how to realise compact, robust and tunable X-ray lasers in the water window, based on ultra-bright electron beams from plasma wakefield accelerators. First, an ultra-low-emittance electron beam is released by a plasma photocathode in a metre-scale plasma wakefield accelerator. By tuning the beam charge, space-charge forces create a balance between beam fields and wakefields that reduces beam energy spread and improves energy stability -- both critical for beam extraction, transport, and focusing into a metre-scale undulator.  Here, the resulting ultra-bright  beams produce wavelength-tunable, coherent, femtosecond-scale photon pulses at ultra-high gain. This regime enables reliable generation of millijoule–gigawatt-class X-ray laser pulses across the water window, offering tunability via the witness beam charge and robustness against variations in plasma wakefield strength. Our findings help democratise access to coherent, high-power, soft X-ray radiation.  
\end{abstract}

\section{Introduction}
Water is transparent to soft X-rays in the 2.3--4.4 nanometre 
wavelength between the oxygen and carbon K-edges. 
Coherent X-ray radiation in this `water-window' is required for high-contrast imaging and spectroscopic investigation of organic molecules in their natural aqueous environment \cite{Pertot2017-HHG-at-water-window,WW_microscopy_kordel}. 
If this radiation is pulsed, with pulse durations in the few femtosecond range, it also allows direct 
observation of electron 
dynamics in molecules and solids at their natural timescale in pump-probe experiments \cite{alexander2024attosecond,li2024attosecond,guo2024experimental}. 
Furthermore, coherent soft X-ray radiation at these wavelengths is increasingly being considered for X-ray lithography for future nanofabrication modalities \cite{bharti_x-ray_2022, maldonado_x-ray_2016}. 
High-harmonic generation (HHG) \cite{krause1992high,l1993high,paul2001observation} and radiofrequency 
powered  linear accelerator (linac) based X-ray free-electron lasers (XFELs) \cite{FLASH_ackermann_2007,first-lasing-lcls-2010} routinely deliver ultra-short radiation pulses at water-window wavelengths with excellent spatial coherence, but each have shortcomings. 

Thanks to commercially available tabletop millijoule, femtosecond-class laser systems, HHG radiation sources are widespread across university-scale laboratories. 
However, the conversion efficiency ($\approx \mathcal{O}(10^{-6})$) \cite{HHG-conversion-Shiner-2009}  drops dramatically with increasing harmonics, resulting in pulse energies of just tens of nanojoules at the desired soft X-ray wavelength \cite{li_attosecond_2020}. 

In contrast, XFELs can generate coherent photon pulses at resonant wavelengths $\lambda_{\mathrm{r}}$ in the soft X-ray range with a million times greater pulse energy. XFELs rely on microbunching of high-quality electron beams at relativistic energies ($\gamma \gg 1$) oscillating in magnetic undulators \cite{huang_review_2007,BrianNatPhoton2010}, where $\gamma \approx W/(m_{\mathrm{e}}c^{2}) + 1$ is the Lorentz factor, $W$ is the electron beam energy, and $m_{\mathrm{e}}c^{2}$ is the rest mass energy of electrons. The conversion efficiency, saturation length, and peak power are crucially determined by the 1D FEL parameter $\rho_\mathrm{1D}\approx \mathcal{O}(10^{-3})$ \cite{huang_review_2007}. 
It can be maximised through the six-dimensional electron beam brightness $B_\mathrm{6D} = 2I_\mathrm{pk}/(\varepsilon_\mathrm{n}^2 0.1\% \sigma_\mathrm{W})$ as $\rho_\mathrm{1D} \propto B_\mathrm{6D}/\gamma^2$ \cite{DiMitri:2014wpa,rosenzweig2024high}, where $I_\mathrm{pk}$ is the peak current, $\varepsilon_\mathrm{n}$ is the normalised emittance, and $\sigma_\mathrm{W}$ is the energy spread of the electron beam. 
At the same time, maximising 6D brightness relaxes the challenging energy spread condition $\sigma_\mathrm{W} \leq \rho_\mathrm{1D}$. 
In turn, minimising the normalised emittance of 
the beam is the golden path to increasing 6D brightness, since $B_\mathrm{6D} \propto \varepsilon_\mathrm{n}^{-2}$. 
Moreover, a low-emittance beam allows the FEL emittance condition $\varepsilon_\mathrm{n} \leq \gamma \lambda_\mathrm{r}/4\rm{\pi}$ to be satisfied at comparatively low energies, which further boosts $\rho_\mathrm{1D}$ and enhances overall FEL efficiency and performance.

 These stringent electron beam quality requirements make 
 XFELs 
 costly machines. 
Firstly, the accelerating gradient in these metre-scale radiofrequency cavities 
is limited to 50--100\,MV\,m$^{-1}$, which necessitates hundreds or even thousands of metres of complex 
 linac infrastructure to reach
 multi-GeV beam energies. 
The limited electric field also restricts the obtainable beam emittance  due to space charge-driven emittance growth during initial electron beam formation at the injector. 
Emittance is further compromised during longitudinal beam compression to kA peak currents with magnetic chicanes. 
Both contribute to a brightness ceiling \cite{DiMitri:2014wpa,rosenzweig2024high}, which in turn necessitates even higher electron beam energy to reach the lasing threshold. 
As a result, both the limited brightness and the need for high beam energy directly constrain gain and conversion efficiency, ultimately requiring long and costly undulator lines to achieve FEL operation.

On the other hand, one cannot simply increase the driving laser pulse power in HHG to obtain larger X-ray pulse powers. 
At higher laser pulse intensities, the magnetic Lorentz force of the laser pulse becomes significant, and plasma electrons are pushed far away from the parent ions via the ponderomotive force. In the process, they obtain keV-scale transverse kinetic energies and are re-attracted towards the laser propagation axis by the collective Coulomb force of the remaining ions, thus forming a plasma wave in the wake of the laser pulse with wavelength  $\lambda_\mathrm{p} \propto n_\mathrm{p}^{-1/2} \approx 30-$\SI{100}{\micro\meter}, where $n_\mathrm{p}$ is the plasma density. 
Some of these `hot' electrons may be injected into the plasma wave, forming high-current beams in the accelerating and focusing phase,
where wakefields reach  $E_{\mathrm{z}}\propto n_\mathrm{p}^{1/2}\approx \,$10--100\,$\text{GV\,m}^{-1}$. 
Therefore, electron beams from such laser wakefield accelerators (LWFA) have been considered as a pathway to enable ultra-compact FELs for a long time. 
After incoherent synchrotron radiation based on spontaneous emission was demonstrated with LWFA-generated electron beams at near-infrared 
\cite{Schlenvoigt2008}, soft X-ray \cite {Fuchs2009} and water-window wavelengths \cite{maier2020water}, recently optimised LWFA-beams enabled self-amplified spontaneous emission (SASE) high-gain FEL activity at 27\,nm 
wavelength and \SI{100}{\nano\joule}-scale 
pulse energy \cite{wang_free-electron_nodate}, as well as seeded high-gain FEL at 270\,nm \cite{LabatHZDRFEL2023}. 
Also, an electron-beam driven plasma wakefield accelerator (PWFA) was used to post-accelerate an electron beam from a linac and trigger lasing at 830 nm \cite{pompili_free-electron_2022}.

However, overcoming emittance and energy thresholds is increasingly difficult for decreasing FEL wavelengths.  
First, in LWFA, the oscillating, transversely hot plasma electrons 
that may be injected into the plasma wave, impose a lower bound on the achievable beam emittance.
Second, the strong accelerating field gradients tend to produce beams with significant energy spread.  
While gradual improvements of electron beam quality from LWFA and PWFA and/or operation at higher electron energy is expected to lead to more reliable FEL operation and higher pulse energy, new concepts will likely be required to advance electron beam emittance and energy spread to a level where water-window wavelengths can be accessed \cite{assmann2020eupraxia,emma2021free,galletti2024prospects,lindstrom2025beam}, and FEL pulse saturation can be reached.  

Plasma photocathodes were conceptually introduced to provide a method for electron beam generation in PWFA that opens up a path to ultra-low emittance, and brightness that is orders of magnitude higher than state-of-the-art \cite{hidding_ultracold_2012,manahan_single-stage_2017}. 
The plasma photocathode shares similarities with the HHG process: like in HHG, a mJ-class near-infrared laser pulse is focused to intensities slightly above a target ionisation level, albeit unlike HHG the goal is to realise this within a beam-driven plasma wakefield.  
As a result, transversely cold electrons are liberated from their parent ions and are then quickly accelerated, focused and compressed by the plasma wakefield, thus rapidly forming a well-defined electron `witness' beam with normalised transverse emittance as low as 10s of nm\,rad \cite{habib_attosecond-angstrom_2023}.

Here, we show how to generate ultra-bright, ultra-low emittance electron beams with charge up to $\sim$100\,pC, while simultaneously minimising energy spread via optimal beam-loading of the plasma wakefield. To achieve this, we employ a strategically integrated  combination of a single plasma photocathode injector and a single dephasing-free PWFA stage to robustly and reliably produce these beams and accelerate them to GeV energies for XFEL applications. We demonstrate how these beams can be transported and utilised to drive a stable, tunable, metre-scale XFEL delivering millijoule-level, gigawatt-class pulses in the water window, operating in a previously unexplored ultra-high-gain FEL regime.
Furthermore, we reveal that tuning the plasma photocathode injector provides a powerful lever for fine control over the XFEL pulse characteristics.

\section{Results}

\subsection{Simultaneous energy spread and emittance optimisation through direct, dephasing-free beam-loading}
\label{sec:PWFA}

In electron-driven PWFA \cite{chen_acceleration_1985, rosenzweig_nonlinear_1987, rosenzweig_experimental_1988}, a relativistic, high-density electron drive beam excites a plasma wave by transversely pushing plasma electrons away from their parent ions through its Lorentz-contracted, unipolar space-charge fields. 
The blowout regime is reached when sufficiently high drive beam density $n_{\mathrm{d}}$ is realised, such that $n_{\mathrm{d}}/n_{\mathrm{p}} \gg 1$. 
The wakefield is created by a pure ion background and possesses linear accelerating and focusing electric fields on the order of 10--100 $\text{GV\,m}^{-1}$, ideal for preserving the emittance of 
witness beams and for rapid acceleration. 
PWFA is dephasing-free, enabling a fixed phase relation between driver and witness beams even over metre-scale acceleration distances \cite{blumenfeld_energy_2007, Litos2014}. 
This is in stark contrast to LWFA, where the accelerated electrons are faster than the drive laser, leading to dephasing and a changing wakefield amplitude at the witness beam position. 
The witness beam represents a local charge and current imbalance in the ion background, and by loading significant charge into the blowout, the witness beam can be used to flatten the local wakefield -- the basis for witness beam energy spread minimisation through beam-loading. 

Inherently dephasing-free PWFA  offers favourable conditions for leveraging beam-loading to minimise energy spreads \cite{Katsouleas:1987yd, tzoufras_beam_2008}.  
The plasma photocathode injection method is particularly suitable for loading the wake \cite{manahan_single-stage_2017}, because the injection rate is decoupled from the driver and wake properties, allowing precise control of charge released and trapped within the wakefield \cite{habib_plasma_2023}. 
Plasma photocathodes are synergistic with the blowout regime, since particularly strong drive beams and accelerating fields are needed to trap cold electrons released by the plasma photocathode. 
Therefore until recently, experimental plasma photocathode demonstration \cite{deng_generation_2019, habib_plasma_2023} was exclusively achievable at the powerful SLAC linac. However, recently at HZDR, a plasma photocathode was realised in PWFA driven by a 500 MeV, nC-class electron beam obtained from LWFA \cite{uferNutterToBePublished}. 
 This hybrid LWFA$\rightarrow$PWFA configuration \cite{HiddingPRL2010PhysRevLett.104.195002short,kurz_demonstration_2021,FoersterPRX2022} removes the experimental bottleneck for advanced PWFA and plasma photocathode realisation, by opening strong-field PWFA to high-power laser laboratories globally.

On the other hand, increased charge release through the plasma photocathode also increases the intra-beam space charge forces during initial witness beam formation, hence increasing the  transverse momentum of the released electrons \cite{manahan_single-stage_2017}.
 This raises the critical question of whether normalised emittances at the nm-rad level can still be preserved when trapping 100 pC-class beams for optimal beam-loading.
 This section addresses this question using the high-fidelity Fourier-Bessel Particle-In-Cell (FBPIC) code \cite{FBPIC}, modelling PWFA across a broad range of fully resolved plasma photocathode working points.

\begin{figure*}[h!]
   \centering
  
    \includegraphics[width = \linewidth]{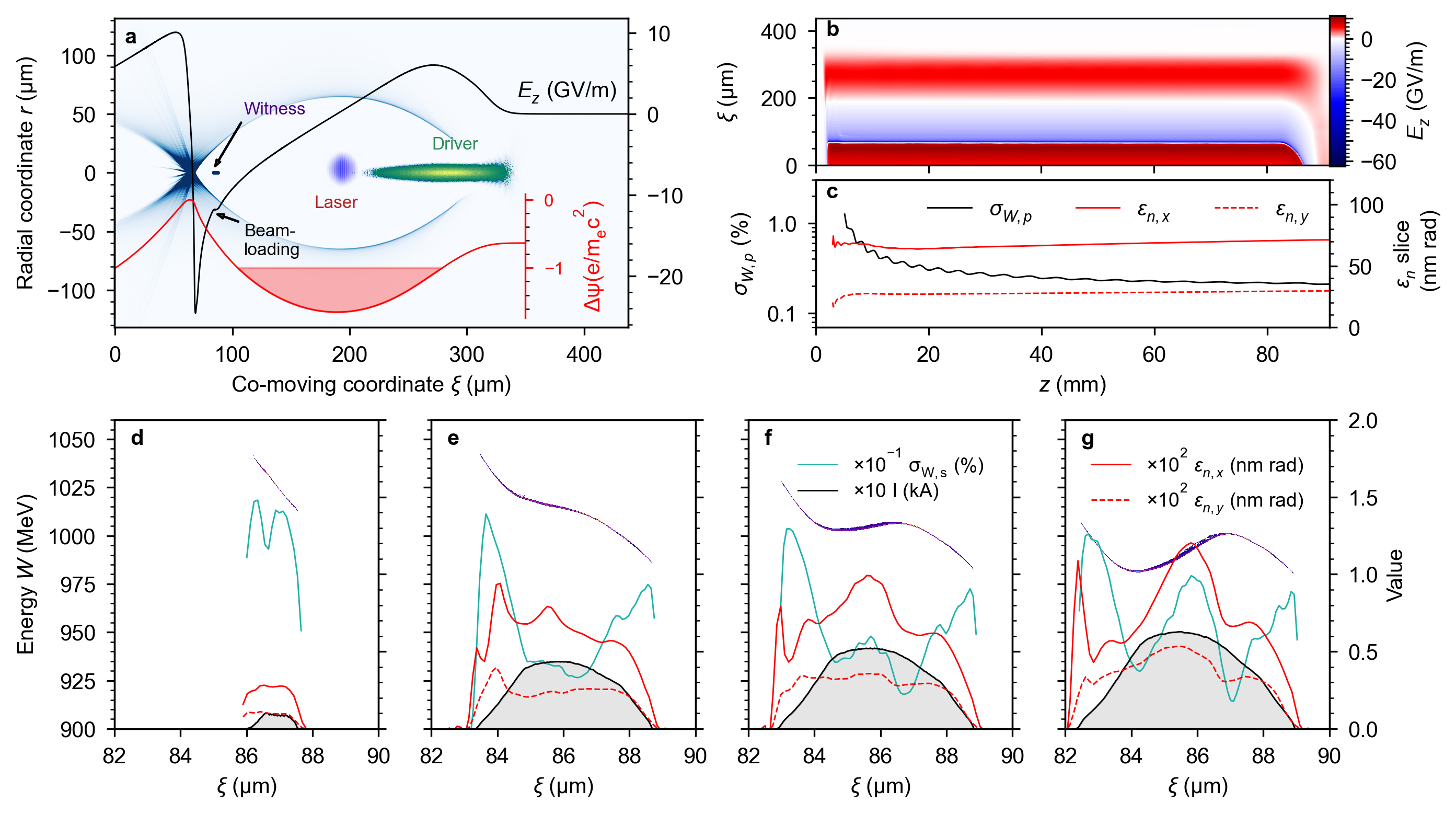}
    \caption{\textbf{Direct plasma photocathode-based beam-loading.} 
\textbf{a}, The drive beam (green) excites a \SI{250}{\micro\meter}-long plasma wave (blue-to-white) in the blowout regime, generating an on-axis wakefield $E_{\mathrm{z}}$ with GV\,m$^{-1}$ amplitude (black solid line). A trapped witness beam with 68\,pC charge (blue), originating from a plasma photocathode and trapped at the co-moving position $\xi = z - ct \approx \SI{86}{\micro\meter}$, flattens the local wakefield via beam-loading. 
\textbf{b}, The evolution of the on-axis wakefield along the propagation direction $z$ shows a nearly constant field, 
ideal for beam-loading. 
\textbf{c}, Witness beam parameter evolution along the PWFA stage shows a reduction in the projected energy spread $\sigma_{\mathrm{W,p}}$ (black), along with preservation of the slice normalised emittance $\varepsilon_{\mathrm{n,s}}$ (red: solid for $x$-plane, dashed for $y$-plane) at the nm-rad level. 
\textbf{d–g}, Final witness beam longitudinal phase spaces at the plasma stage exit, overlaid with current $I$ (solid black), slice relative energy spread $\sigma_{\mathrm{W,s}}$ (solid cyan), and slice normalised emittance profiles $\varepsilon_{\mathrm{n,(x,y),s}}$ (red solid and dashed lines), are shown for \textbf{d}, low charge (\SI{3}{pC}), \textbf{e}, underloaded (\SI{52}{pC}), \textbf{f}, optimally loaded (\SI{68}{pC}), and \textbf{g}, overloaded (\SI{89}{pC}) witness beams. Further details are given in Extended Data Fig. 1. 
    }
   \label{fig1}
\end{figure*}

We operate the PWFA stage at comparatively 
low plasma density $n_{\mathrm{p}}\approx 1.78 \times 10^{16} \,\mathrm{cm^{-3}}$ of singly-ionised helium, resulting in a large blowout structure with plasma wavelength of $\lambda_\mathrm{p} \approx $ \SI{250}{\micro\meter}  (see Fig. \ref{fig1}a, blue colour coding). 
Employing such  large blowouts has several advantages. 
First, it significantly relaxes the spatio-temporal alignment precision requirements of the plasma photocathode injector laser pulse \cite{habib_plasma_2023}. 
Second, it minimises the residual energy spread of the witness beam due to the reduced accelerating gradient at the witness beam trapping position (black solid line) \cite{manahan_single-stage_2017}, enabling low relative slice energy spread  at comparatively low beam energies. 
Finally, the peak current requirement for optimal beam-loading by the witness beam is lowered, thereby reducing space charge-driven emittance growth during its formation.  

The electron driver beam has charge $Q \approx 1.26\,\mathrm{nC}$ at \SI{1}{GeV} energy, with a root mean square (rms) longitudinal length $\sigma_{\mathrm{z,d}} \approx $ \SI{25}{\micro\meter} and radius $\sigma_{\mathrm{r,d}}\approx $ \SI{2.5}{\micro\meter}, corresponding to a beam density (see Fig. \ref{fig1}a, green colour coding) of $n_{\mathrm{b}}\approx (Q/e)/(2\pi)^{3/2} \sigma_{\mathrm{r,d}}^2\sigma_{\mathrm{z,d}} \approx 3.2 \times10^{18}$ cm$^{-3}$, where $e$ is the elementary charge.
The high drive beam density 
($n_{\mathrm{d}}/n_{\mathrm{p}} \approx180$) thus excites 
a blowout with deep electrostatic trapping potential of $\Delta \Psi <-1 $ (Fig. \ref{fig1}a, red solid line) by a comfortable  margin, 
satisfying the 
requirement 
for capturing electrons released from the plasma photocathode \cite{hidding_ultracold_2012,habib_plasma_2023}.

Then, a fully resolved 
$\lambda=$ \SI{800}{nm} wavelength, sub-mJ class, collinear plasma photocathode laser pulse (shown schematically on top of the simulation snapshot) with full-width half maximum (FWHM) pulse duration $\tau$ = \SI{60}{fs} is focused to an rms spot size $w_0 =$ \SI{7}{\micro\meter}  at $z\approx$ \SI{2}{\milli\meter} in the laboratory frame. 
For an initial working point, the laser pulse energy is set to  reach an intensity of $I_0 \approx 7.4\times10^{15}\,\mathrm{W\,cm^{-2}}$, corresponding to a normalised laser amplitude 
 $a_0 \approx 0.85 \times 10^{-9} \, \lambda \, [\mathrm{\mu m}] \times (I_0 \, [\mathrm{W\,cm^{-2}}])^{1/2}
 \approx 0.0600$. This is sufficient to overcome the tunnel ionisation threshold of the background He$^+$ ions of \SI{54.4}{eV} in a confined region of the laser focus (see Methods \ref{sec:methods_plasma_stage}, ionisation model details), and the plasma photocathode  releases $Q_{\mathrm{w}}\approx$ \SI{3}{pC} charge  
 in the centre of the plasma wave. 
 
These initially non-relativistic electrons are rapidly compressed both transversely and longitudinally during the trapping process. They are then accelerated by the wakefield as a witness beam co-propagating phase-constantly at the back of the blowout. 
After rapid acceleration at $E_\mathrm{z}\approx 12 $ $\text{GV\,m}^{-1}$, this low charge witness beam  reaches $W \approx $ \SI{1.023}{GeV} after $z\approx 91$\,mm at the exit of the PWFA stage. 
The uniformity of the longitudinal acceleration is visualised in Fig. \ref{fig1}b.   
This beam, analysed in Fig. \ref{fig1}d, exhibits ultra-low slice normalised emittance $\varepsilon_{\mathrm{n,x,s}} \approx$ \SI{27}{\nano\meter \radian} and  $\varepsilon_{\mathrm{n,y,s}} \approx$ \SI{10}{\nano\meter \radian}, which would support lasing even in the hard X-ray range. 
The emittance is significantly larger in $x$ than in $y$ due to a non-negligible thermal emittance contribution \cite{ThermalEmittancePhysRevSTAB.17.101301Schroeder2014} from the laser pulse, which is linearly polarised in $x$-direction. 
However, despite being ultra-short ($\approx$ \SI{1}{fs} rms), the witness beam also exhibits significant energy chirp, which is problematic for beam quality preservation during transport, and eventually lasing in the undulator. 

Now, the tunability of the plasma photocathode  is exploited for direct beam-loading in the phase-constant PWFA. 
By tuning the laser intensity -- an experimentally straightforward parameter to change through laser pulse energy variation -- the plasma photocathode releases varying amounts of helium electrons. 

In Fig. \ref{fig1}e-g, we gradually increase the normalised laser amplitude from 0.0850 to 0.0960. 
This increases the tunnelling ionisation volume and 
ionisation front movement in the co-moving frame. 
As a result of this and ensuing transient space-charge driven elongation, the charge, length, peak current, and emittance of the formed witness beams are increased -- while 
chirp and slice energy spread are reduced. 
The beam charges are now more than an order of magnitude higher, ranging from 52--89\,pC, 
 which produces significantly larger peak currents of $I_{\mathrm{pk}} \approx$ 4--6\,kA. 
These peak currents are sufficiently high to modify the local wakefield at the witness beam trapping position, and shape the longitudinal phase space of the witness beam. 
A gradual transition from positively chirped (Fig. \ref{fig1}e), to optimally loaded (Fig. \ref{fig1}f) and overloaded witness beams (Fig. \ref{fig1}g) is achieved, simply by changing the plasma photocathode intensity. 
The projected energy spread decreases due to beam-loading, and the slice energy spread also decreases on the scale of the FEL cooperation length ($L_\mathrm{c} \sim 115$\,nm, see Methods \ref{sec:FELMethodes}) because of this phase space flattening. 
Equally important for XFEL applications, the average normalised slice emittances are still exceptionally low, despite the substantially higher charge and peak current, ranging from $\varepsilon_{\mathrm{n,y,s}} \approx$ \SI{30}{nm\,rad} to $\varepsilon_{\mathrm{n,x,s}} \approx$ \SI{70}{nm\,rad}.
This trade-off between current, emittance and energy spread is crucial for the FEL performance: it 
converts the 
excess slice emittance budget into reduced slice energy spread and chirp, resulting in beam slices that lie safely under both the lasing emittance and energy thresholds (see Section \ref{sec:FEL}).   
The multi-kA slice currents 
further increase beam brightness.

Operation in the dephasing-free strong-field PWFA blowout regime, with its constant, linear focusing fields, enables complete emittance preservation and quasi-static acceleration after the initial witness beam formation. 
Figure \ref{fig1}c, corresponding to the witness beam with optimal beam-loading and charge $Q_\mathrm{w} \approx 68$\,pC displayed in Fig. \ref{fig1}f and Fig. \ref{fig1}a, highlights that normalised slice emittances remain constant throughout the acceleration process. 
The projected relative energy spread, in contrast, reduces to $\sigma_{\mathrm{W,p}} \approx  \SI{0.21}{\%}$ by the end of the acceleration, while the slice energy spread reaches $<0.05\%$ (see Methods \ref{sec:methods_plasma_stage}).

\begin{figure*}[!h]
    \centering
    \includegraphics[width = \textwidth]{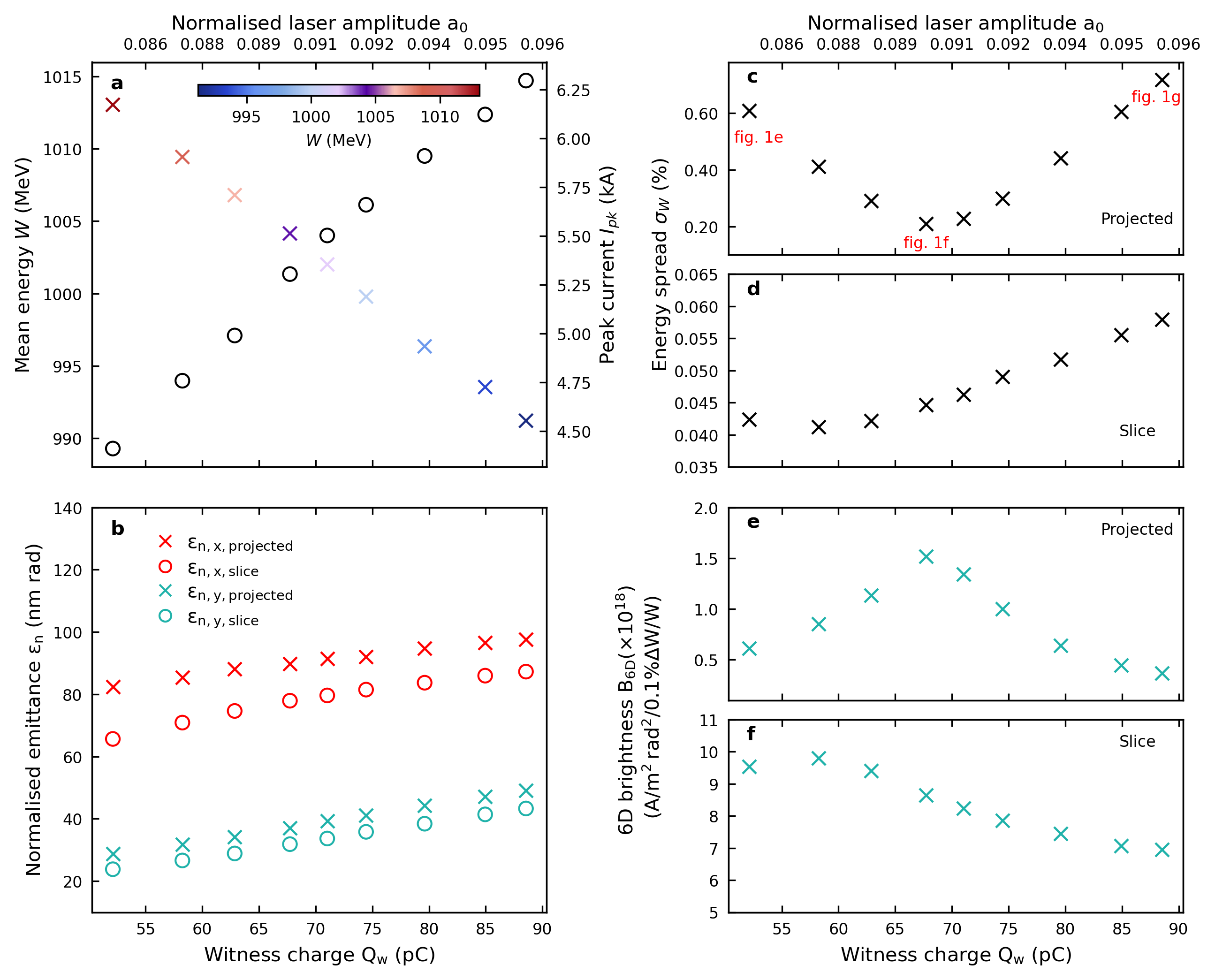}
    \caption{\textbf{Scan of working points in plasma photocathode PWFA stage.} 
\textbf{a}, Witness beam energy $W$ as a function of witness charge $Q_{\mathrm{w}}$ (bottom $x$-axis), released by the corresponding normalised laser amplitude $a_0$ (top $x$-axis) of the plasma photocathode laser pulse. The peak current (right $y$-axis) scales linearly with the witness beam charge, enabling fine tuning of the local wakefield at the witness position via beam-loading. 
\textbf{b}, Corresponding witness beam projected ($\times$) and slice ($\mathrm{o}$) normalised emittances in the $x$ (red) and $y$ (cyan) planes also scale linearly with the witness beam charge and peak current. 
\textbf{c}, Projected energy spread is minimised at $Q_{\mathrm{w}} = 68$\,pC, reaching $\sigma_{\mathrm{W,p}} \approx 0.21\,\%$, corresponding to the purple data point in \textbf{a}. The underloaded, optimally loaded, and overloaded witness beams from Fig.~\ref{fig1} are labelled in red. 
\textbf{d}, Slice energy spread increases with $Q_{\mathrm{w}}$ (peak current) due to a non-negligible contribution from the nonlinear energy chirp and ranges from $\sigma_{\mathrm{W,s}} \approx 0.041\text{--}0.058\,\%$. 
\textbf{e}, The projected 6D brightness peaks at $B_{\mathrm{6D,p}} \approx 1.5 \times 10^{18}\,\mathrm{A}\mathrm{m}^{-2} \mathrm{rad}^{-2}0.1\%\sigma_{\mathrm{W,s}}^{-1}$ for the optimally loaded witness beam. 
\textbf{f}, Slice brightness is an order of magnitude higher, ranging from $B_{\mathrm{6D}}\approx 7\text{--}10 \times 10^{18}\,\mathrm{A}\mathrm{m}^{-2} \mathrm{rad}^{-2}0.1\%\sigma_{\mathrm{W,s}}^{-1}$. 
    }\label{fig2}
\end{figure*}

The variation of normalised laser amplitude (used in Fig. \ref{fig1}e-g) amounts to $\sim1\,\%$ between each beam, such that each laser setting represents an 
individual witness beam working point that may be selected. 
Figure \ref{fig2} reveals the highly systematic impact of $a_0$ variation on resulting witness beam parameters through beam-loading. 
Figure \ref{fig2}a shows that the peak current increases approximately linearly, while the mean energy decreases approximately linearly with $a_0$. 
Meanwhile, Fig. \ref{fig2}b shows that  projected (slice) normalised emittances increase approximately linearly, albeit only by a few nm\,rad, despite a charge increase of almost $100\%$. 
Figure \ref{fig2}c and d reveal that while the relative average slice energy spread increases with charge, it remains $<0.055$\% for all witness charges due to the sub-0.5\,MeV residual energy spread. 
The projected energy spread changes parabolically, reaching the minimum value of \SI{0.21}{\%} for the optimum beam-loading 
working point corresponding to Fig. \ref{fig1}f. 

Because the change of normalised emittances and currents across the working points is comparatively small, the 6D brightness 
(see Fig. \ref{fig2}e \& f) is dominated mainly by the relative energy spread. 
The projected 6D brightness peaks at $B_{\mathrm{6D}}\approx 1.5 \times 10^{18}\,\mathrm{A}\mathrm{m}^{-2} \mathrm{rad}^{-2}0.1\%\sigma_{\mathrm{W,p}}^{-1}$, exactly at the optimum beam-loaded position, and decreases 
for higher and lower charge beams. 
The slice 6D brightness across the working points is  $B_{\mathrm{6D}}\approx 7\text{--}10 \times 10^{18}\,\mathrm{A}\mathrm{m}^{-2} \mathrm{rad}^{-2}0.1\%\sigma_{\mathrm{W,s}}^{-1}$, one order of magnitude higher compared to the projected values. 
Notably, all ultra-bright beams exhibit slice energy and emittances well below the required FEL thresholds. 
This enables one to conveniently tune the working point of the witness beam in the PWFA stage by varying just a single parameter -- the intensity of the plasma photocathode laser pulse -- without compromising the suitability for FEL application. 
Experimentally, this is straightforward and allows systematic control of electron beam properties. 
In contrast, achieving comparable tunability in conventional linacs would be considerably more complex due to the interdependency of the beamline and linac elements. 

\subsection{Beam transport}\label{sec:beamline}
Production of multi-kA, few-femtosecond electron beams with nanometre-radian scale emittances and few-100-keV-scale residual energy spread at 1 GeV energy offers huge potential for ultra-high gain and robust XFEL operation.
However, with these beams one enters unknown territory, since on the one hand the corresponding 6D brightness is many orders of magnitude larger than state-of-the-art, and on the other hand they experience strongly transverse focusing forces within the plasma wakefield, leading to small 
beta functions ($\beta\leq$ \SI{1}{\milli\meter}) $-$ the equivalent of the Rayleigh length of light pulses -- when exiting the PWFA. 
Consequently, they expand with significant divergence into the vacuum, which requires quick and strong capturing and refocusing into subsequent transport line elements. 
In this process, chromatic aberration may lead to significant emittance growth \cite{Floettmann2003, Mehrling2012, migliorati_intrinsic_2013,andre2018control,PhysRevX.10.031039} e.g. on micrometre-radian scale $-$ and the impact of such beam quality degradation on ultra-low emittance beams \cite{manahan_single-stage_2017,habib_attosecond-angstrom_2023} would be catastrophic. 
Furthermore, a $\simeq$\SI{1}{\%} electron beam energy jitter already complicates beam capturing and matching \cite{Carl2016achromat}, leading to chromatic mismatch in the beam transport line. 

This is where the reduction in both projected and slice energy spread due to beam-loading comes into play, now proving to be extremely beneficial for achieving preservation of high beam quality during transport. 
We designed a capture and transport beamline optimised for 
the optimally beam-loaded witness beam shown in Fig. \ref{fig1}f. 
The transport beamline begins with 
four permanent magnet quadrupoles (PMQs) that capture the rapidly expanding witness beam in the drift space after the plasma stage without charge loss. 
Such a PMQ quartet increases the momentum acceptance, helping to minimise emittance growth caused by the remaining nonlinear curvature in the longitudinal phase space of the witness beam. At the same time, it collimates the beam such that the Twiss parameters satisfy $\alpha_{\mathrm{x,y}} \approx 0$ (see Extended Data Fig. 2) 
at the entrance of the subsequent electromagnetic quadrupole (EMQ) triplet.
Then, the EMQ triplet focuses the witness beam downstream, positioning the beam waist approximately at the centre of the undulator. 

\begin{figure*}
    \centering
    \includegraphics[width = \textwidth]{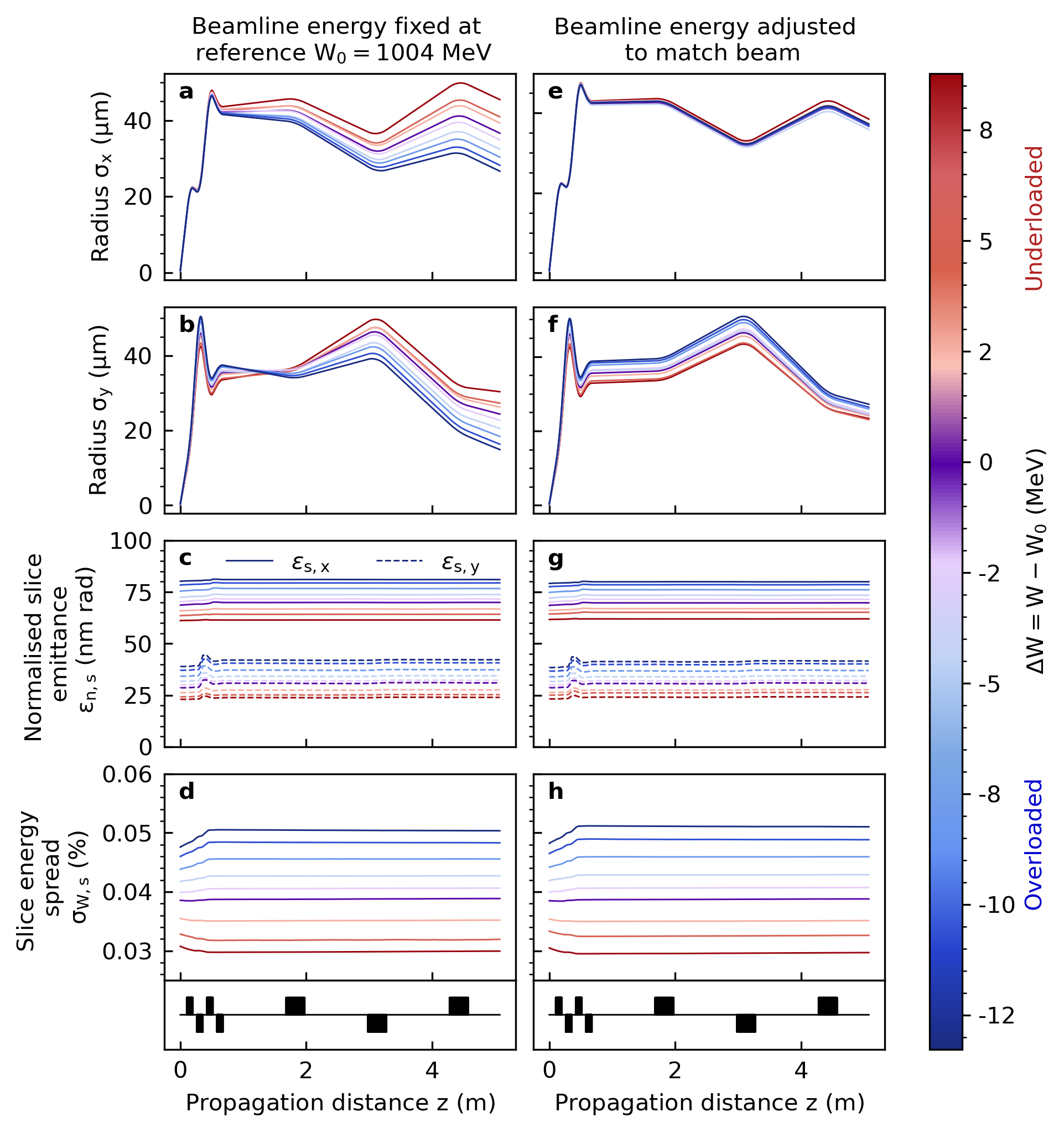}
   \caption{\textbf{Beam transport line simulation.} Beam transport line (bottom plot, black blocks) featuring a PMQ quartet for initial beam capture and an EMQ triplet for matching to the undulator. The evolution of witness beam parameters as a function of beamline distance $z$ is shown for two configurations: (left column) fixed beamline energy at the reference beam energy $W_0 \approx 1004\,\mathrm{MeV}$, and (right column) beamline energy adjusted to each individual witness beam energy. From top to bottom, panels \textbf{a–f} display the transverse beam radii, while panels \textbf{c, g} show the slice normalised emittances in the $x$- and $y$-plane, and \textbf{d, h} present the evolution of slice energy spread. Colour coding indicates the absolute energy deviation $\Delta W = W - W_0$ of each witness beam relative to the reference beam energy.
   The slice normalised emittance remains nearly constant across all configurations and witness beams, indicating nm-rad level beam quality preservation. 
   }
    \label{fig3}
\end{figure*}

We simulated this scenario with the tracking code Elegant\cite{borland_elegant_2000} (see Methods \ref{sec:TransportMethodes}). Initially, we set the beamline to the reference energy ($W_0 \approx$ \SI{1004}{MeV}, $\gamma \approx$ \SI{1964}{}) in the beam transport simulation (see Methods \ref{sec:TransportMethodes}) and track all witness beams shown in Fig. \ref{fig2} through the beamline to the undulator entrance.
Figures \ref{fig3}a-b show that witness beams at the reference energy (purple lines) are collimated  ($\alpha_{\mathrm{x,y}}\approx 0$, see Extended Data Fig. 2) 
at the entrance of the EMQ triplet in both the $x$- and $y$-plane.
By contrast, beams with higher (red lines) or lower (blue lines) energies are under-focused or over-focused, respectively, and enter the EMQ triplet with divergence and different beam sizes. 
These beam envelope variations propagate downstream, leading to an even larger beam size spread at the undulator entrance.

The beam energy, along with the variations in projected and slice emittances and energy spreads, plays a central role in determining the beam size and focusing performance along the transport line and at the undulator entrance. 
As outlined in Section~\ref{sec:PWFA}, each beam configuration corresponds to a distinct working point in the PWFA stage. 
These working points can be 
chosen by fine-tuning the vector potential $a_0$ of the plasma photocathode pulse, correlating nearly linearly with final electron beam energy. 
Accordingly, the beamline can be optimised to these working points and the associated range of witness beam energies using gradient-adjustable PMQs and EMQs~\cite{andre2018control,LabatHZDRFEL2023}. 
The result is a significantly reduced spread in beam sizes at the entrance to the undulator (see Fig. \ref{fig3}e,f vs. \ref{fig3}a,b), with minimal change to the obtained slice emittance and energy spread evolution and values (see Fig. \ref{fig3}g,h vs. \ref{fig3}c,d). 

Most importantly for the FEL interaction, Figs. \ref{fig3}c and \ref{fig3}g reveal that the slice emittances -- in the range of $\varepsilon_{\mathrm{n,y,s}} \approx$ 25--30\,nm\,rad and $\varepsilon_{\mathrm{n,x,s}} \approx$ 67--87\,nm\,rad -- remain nearly constant throughout the entire transport line. More specifically, the underloaded, higher-energy (red) beams exhibit and maintain smaller slice emittances than the overloaded, lower-energy (blue) beams in both transverse planes (cf. Figs. \ref{fig1} and \ref{fig2}). 
Overall, the initial slice emittance at the tens of nm-rad level grows negligibly along the beam transport line, in stark contrast to the \SI{1}{\micro\meter}-rad-scale emittance growth usually encountered for electron beams produced
in plasma-based accelerators \cite{Mehrling2012, migliorati_intrinsic_2013, andre2018control}.

This preservation of low emittance is enabled by the combination of an initial sub-$0.05\,\%$ slice energy spread and its preservation at the $0.01\,\%$-level (see Figs. \ref{fig3}d and \ref{fig3}h), which effectively minimizes chromatic aberrations and phase-space dilution of the individual emittance slices of the witness beams. We note that the minor change of slice energy spreads after the PMQ quartet is due to non-negligible dispersion, resulting in slight longitudinal elongation or compression of the witness beam when operating in underloaded or overloaded beam-loading regimes in the PWFA stage, respectively.

Furthermore, all other relevant beam slice properties across different working points remain invariant during transport, ensuring full preservation of the slice 6D brightness at the $B_{\mathrm{6D}} \approx 10^{19}\,\mathrm{A}\mathrm{m}^{-2} \mathrm{rad}^{-2}0.1\%\sigma_{\mathrm{W,s}}^{-1}$-level.
To our knowledge, this is the first scenario that shows robust production and transport of multi-kA, few-femtosecond electron beams with nanometre-radian-scale emittances and few-100-keV-scale energy spreads across a wide range of beam-loaded PWFA configurations. 
The core finding -- that ultrahigh brightness is consistently preserved, across all working points -- is an enabling feature for robust operation and reliable realisation of ultra-high-gain XFELs under realistic conditions.  

\clearpage
\subsection{Free-electron laser stage}\label{sec:FEL}
To assess the capability of these ultra-bright electron beams to drive a soft X-ray FEL, and to evaluate the impact of the beam transport line on FEL performance, we performed time-dependent, 3D 
FEL simulations using the unaveraged FEL code Puffin \cite{campbell_puffin_2012,Puffin-update-2018} (see Methods~\ref{sec:FELMethodes}). 
These simulations were carried out for both the static and energy-adjusted beamline scenarios discussed in Fig. \ref{fig3}.
   
We use a single \SI{10}{\meter} long undulator section, 
plane-polarised in the $y$-direction with 
$\lambda_\mathrm{u} =$ \SI{15}{\mm} period and peak field strength $B =$ \SI{0.42}{T}. 
The corresponding resonant wavelength $\lambda_{\mathrm{r}}=(\lambda_{\mathrm{u}}/2\gamma^{2})(1+K^{2}/2)$, where $\lambda_{\mathrm{u}}$ is the undulator period,
and $K\propto B\lambda_{\mathrm{u}}$ is the undulator parameter, ranges from $\lambda_\mathrm{r} =$ 2.2--2.3\,nm for the range of electron beam energies.
At these wavelengths, the emittance criterion $\varepsilon_\mathrm{n,x,s} \approx 67\text{--}87$\,nm\,rad $\,\leq \gamma\lambda_\mathrm{r}/4\mathrm{\pi} \approx 357$\,nm\,rad shows that despite the beam-loading, the electron beams offer an 
excess slice emittance budget for lasing even at this relatively low electron energy. 
While the slice emittance in $y$ is significantly better than in $x$, we deliberately choose the $x$-plane for lasing to emphasise robustness: if successful lasing is demonstrated under the less favourable transverse conditions, lasing in $y$ can be even more confidently expected.  

\begin{figure*}
    \centering
    \includegraphics[width = \textwidth]{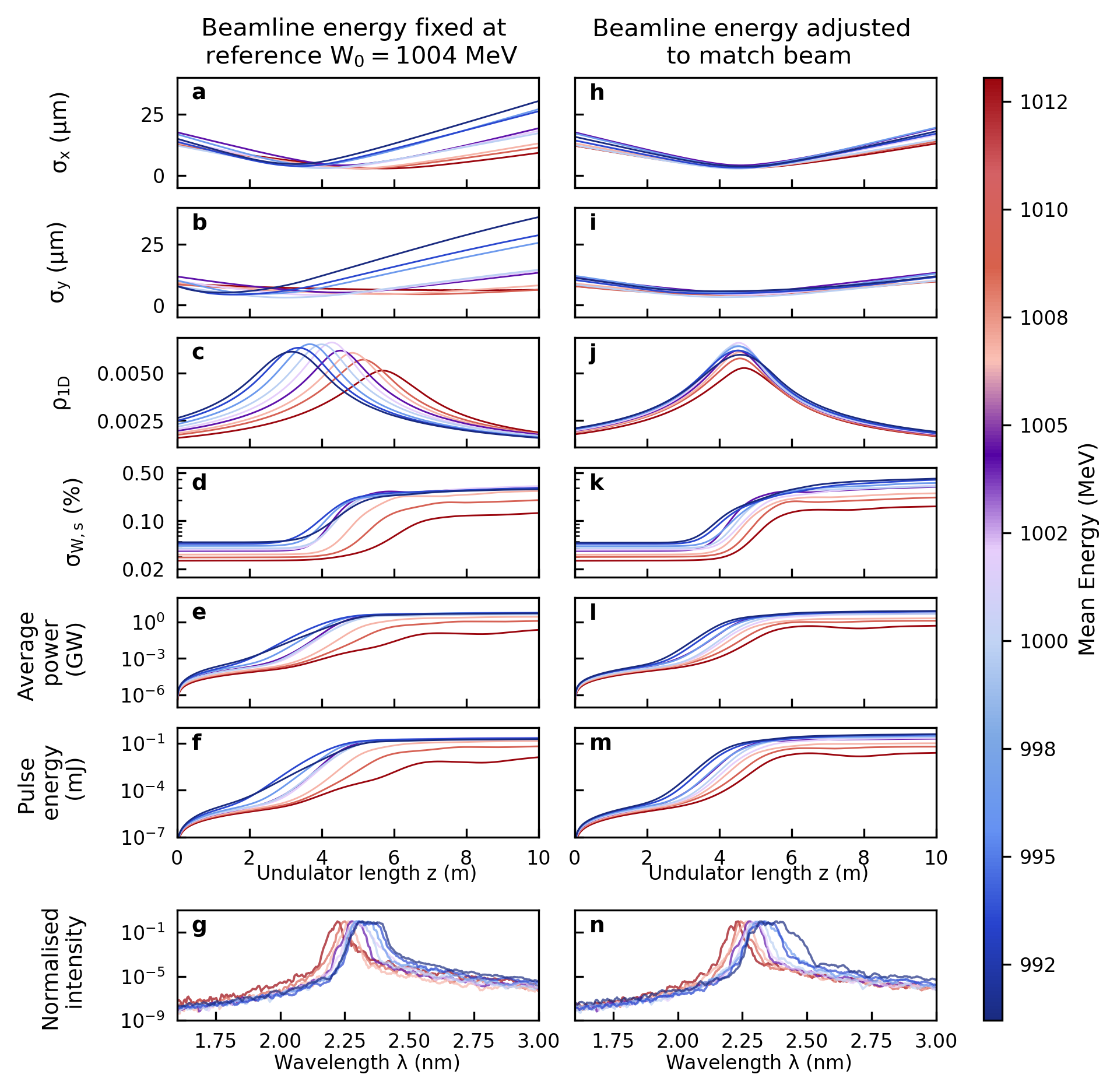}
    \caption{\textbf{Ultra-high-gain water-window free-electron lasing.} 
Simulated performance of 
water-window FEL 
across the identified working points. 
Left column: fixed energy beamline, right column: energy adjusted beamline configuration. 
\textbf{a, h}, Transverse beam radius $\sigma_{\mathrm{x}}$ and \textbf{b, i}, $\sigma_{\mathrm{y}}$ along the undulator.  
\textbf{c, j}, One-dimensional FEL coupling parameter $\rho_{\mathrm{1D}}$ and  
\textbf{d, k}, average slice energy spread $\sigma_{\mathrm{W,s}}$. 
\textbf{e, l}, Evolution of average output power and  
\textbf{f, m}, corresponding pulse energy gain along the undulator.
\textbf{g, n}, Normalised spectral intensity around the fundamental FEL wavelength. The colour scale encodes the mean energy of each witness beam, highlighting robust and reliable FEL interaction across the beam-loading and beam focusing regimes.} 

    \label{fig4}
\end{figure*}

The Xie parametrisation \cite{ming_xie_design_1995} allows the FEL output power and gain length to be estimated depending on undulator and beam slice parameters, and we evaluated and optimized the XFEL performance using this method (see Extended Data Fig. 3).
The 1D FEL parameter is estimated as $\rho_\mathrm{{1D}}=\left[\left(\frac{I_\mathrm{{pk}}}{I_\mathrm{A}}\right)\left(\frac{\lambda_\mathrm{u} A_\mathrm{w}}{2 \pi \sigma_{\mathrm{x}}}\right)^2\left(\frac{1}{2 \gamma}\right)^3\right]^{1 / 3}
\approx 1.6\text{--}2.6\times 10^{-3}$, where $I_\mathrm{A}$ is the Alfvén current, 
and $A_\mathrm{w}$ depends only on the undulator parameters as defined in Ref. \cite{ming_xie_design_1995}.  
Thus the beams satisfy the (slice) energy spread criterion $\sigma_\mathrm{W,s} \approx 4.5\times 10^{-4} \leq \rho_\mathrm{{1D}}$ with room for an order of magnitude increase. The estimated  
 1D power gain length $L_\mathrm{{g_0}} = \lambda_\mathrm{u}/(4\mathrm{\pi}\sqrt{3}\rho_\mathrm{{1D}}) \approx 20$\,cm is one order of magnitude shorter compared to linac-powered XFELs. \cite{campbell_puffin_2012}, a 3D unaveraged FEL code. 
Figure \ref{fig4} shows the evolution of the 
beam radii $\sigma_{\mathrm{x,y}}$ in the oscillation plane $x$ as well as in $y$, the FEL parameter $\rho_\mathrm{1D}$, the average slice energy spread $\sigma_{\mathrm{W,s}}$, the
average power gain, pulse energy, and the normalised spectral intensity. 
The ultra-low normalised emittance $\varepsilon_\mathrm{n}$ allows for small bunch radii $\sigma_x = \sqrt{\frac{\beta^{*} \varepsilon_\mathrm{n}}{\gamma}} \approx$ \SI{3}{\micro\meter} to be realised at comparatively long beta-functions $\beta^{*} \approx$~\SI{1}{\meter} at waist, compared to beams with larger emittance. 
Combined with the ultra-short gain length, no external focusing optics are required in the undulator to reach saturation. 
This strongly alleviates the complexity of beam matching associated with typical quadrupole FODO focusing optics in conventional XFELs. 

For the static beamline scenario (Fig.~\ref{fig4}, left panels), the witness beam at the reference energy ($W_0 \approx$ \SI{1004}{MeV}, purple line) reaches its waist at the centre of the undulator (see Fig.~\ref{fig4}a and \ref{fig4}b), where the FEL parameter $\rho_\mathrm{1D}$ reaches its maximum (cf. Figs.~\ref{fig4}c). 
As a result of the different beam energies and focusing of the transport line in the static scenario (cf. Fig. \ref{fig3}), the electron beams with lower mean energy (blue) reach their minimum $\sigma_\mathrm{x}$ and $\sigma_\mathrm{y}$ earlier in the undulator than beams with higher electron energies (red), whereas the energy-adjusted beamline (Fig.~\ref{fig4}, right panels) compensates for different beam energies at the different working points so that all electron beams reach similar minimum beam size at a similar undulator position (see Figs. \ref{fig4}a-b,h-i). 
 
The evolution of the 1D FEL parameter (see Figs. \ref{fig4}c,j) along the undulator underscores its dependence on the radius of the electron beam $\rho_\mathrm{1D}\propto \sigma_{\mathrm{x}}^{-2/3}$. 
Hence, the FEL parameter increases during focusing, reaching $\rho_\mathrm{1D} > 0.005$ at the beam waist, and subsequently decreases as the beam diverges downstream (cf. Figs. \ref{fig4}c,j). 
The average relative slice energy spread shown in Fig. \ref{fig4}d,k is initially much smaller than the FEL parameter, and increases during microbunching and energy exchange until $\sigma_\mathrm{W} \leq \rho_\mathrm{{1D}}$ is violated and further power gain is terminated. 
Hence, the values of $\rho_\mathrm{1D}$ fundamentally dictate the final energy extraction efficiency of the XFEL. 
We note that these comparatively large $\rho_\mathrm{{1D}}$ values are %
linked to the 6D brightness via $\rho_\mathrm{1D}\simeq (\gamma^{6}/3 )( \lambda_{\mathrm{r}}/2\pi)^4 B_\mathrm{6D}$ \cite{DiMitri:2014wpa,rosenzweig2024high}. 
Therefore, thanks to the ultrahigh slice 6D brightness on the order of $B_{\mathrm{6D}} \approx 10^{19}\,\mathrm{A}\mathrm{m}^{-2} \mathrm{rad}^{-2}0.1\%\sigma_{\mathrm{W,s}}^{-1}$ already achieved at moderate electron energy, and the strong $\gamma^{-2}$ dependence of the resonant wavelength $\lambda_{\mathrm{r}}$, the FEL parameter  $\rho_\mathrm{1D}$ is very high for all beams considered, despite the short resonant wavelength. 
The energy-adjusted beamline scenario that focuses the beams into the same undulator location also homogenises the $\rho_\mathrm{1D}$ evolution, and yields optimisation of the extraction efficiency. 
Fig. \ref{fig4}e,l shows the average power gain. 
Due to the very low relative slice energy spread and emittance, and ultra-high brightness, microbunching builds up rapidly, and lasing via the SASE mechanism sets in quickly. 
The 3D gain length is extremely short and ranges from only 20--60\,cm for all beams, 
which is near the estimated ideal 1D gain length of $L_\mathrm{{g_0}}\approx 20$\,cm. 
The corresponding saturation length is likewise extremely short, amounting to only $L_\mathrm{sat}\approx$ 4.8--7.0\,m. In terms of X-ray pulse energy, up to 0.4\,mJ is extracted from the electron beam (Fig. \ref{fig4}f,m) and delivered in $\approx 5\,$fs FWHM duration pulses (Extended Data Fig. 4), corresponding to high  average pulse power pulse levels, ranging from 0.3--8.5\,GW.  
We anticipate that the 
large emittance and energy spread budget in combination with the energy chirp control capability would be suitable for undulator and energy tapering techniques routinely employed in conventional SASE XFELs \cite{BrianNatPhoton2010}. This may enable the production of  X-ray pulses with even higher pulse energies. 

Our results reveal and define a distinct \textit{ultra-high-gain} FEL regime, 
characterised by ultra-short gain and saturation length in combination with high energy extraction efficiency and GW-level pulse power, uniquely enabled by generation, transport and application of beams with ultra-high brightness in six-dimensional phase space.
A quantitative definition of the ultra-high-gain regime can be introduced by comparing the nominal 1D with the 3D gain length that includes energy spread and emittance degradation effects, as $L_\mathrm{g}/L_\mathrm{g_0} \lessapprox 1.5$ (see Extended Data Fig. 3).   

We note that maintaining longitudinal overlap between the radiation pulse and ultra-short electron beams (FWHM $\tau_{\mathrm{w}}\approx$\, 12 fs) typically becomes increasingly challenging at longer lasing wavelength due to the total slippage time scaling, $S_{\mathrm{sat}}=L_\mathrm{sat}\lambda_{\mathrm{r}}/(\lambda_{\mathrm{u}}c)$.
The  $S_{\mathrm{sat}}\gg \tau_{\mathrm{w}}$ regime can even prevent saturation of the FEL \cite{PhysRevA.40.4467-Bonifacio-Superrad}, because then the radiation pulse will outrun the electron beam, with detrimental implications for the energy extraction efficiency and radiation pulse coherence. 
However, in our scenario, the short saturation length of the ultra-high-gain regime yields short slippage time $S_{\mathrm{sat}}\approx $\, 2.4--4.0\,fs and alleviates the impact of slippage by safely operating in the $S_{\mathrm{sat}}\ll \tau_{\mathrm{w}}$ regime. 
As a result,  the FEL reaches saturation before the radiation pulse outruns the ultra-short electron beam, producing ultra-short, high-power radiation pulses in the water window regime.  

In both beamline scenarios, there is a direct correlation between electron beam energy and peak current on the one hand, and FEL power and energy on the other. Electron beams with lower mean energy carry higher charge and peak current. While each electron on average has less energy, the total electron beam power $P_\mathrm{B}$ is larger for lower electron energy (blue) than for higher electron energy (red) beams. 
At the same time, the FEL parameter, though evolving, is typically lower for higher electron beam energies (red), as shown in Fig. \ref{fig4}c,j. Therefore, the total FEL pulse energy and saturation power is larger for lower electron energy beams (see Fig. \ref{fig4}e,i and f,m), and the saturation length is shorter.      
The intensity distribution is shown in Fig. \ref{fig4}g, confirming that the 
higher energy beams lase at  slightly shorter wavelengths than the lower energy beams. 
Since the individual electron beam working points can be selected by adjusting the plasma photocathode laser intensity, different pulse power levels as well as wavelengths can be chosen simply by tuning the plasma photocathode. 

The energy-adjusted beamline fully leverages the performance capability of the beams. Therefore peak pulse energies and powers can be higher, and there is significantly less variation not only in the beam radii evolution but also the derived FEL quantities, including the saturation length and power. 

\subsection{Robustness and tunability of the XFEL}
\label{robustness}

The large emittance and energy spread budget of produced witness electron beams, and the observed ultra-high gain, implies robustness and tunability. 
With regard to robustness, a particularly relevant parameter is the electron beam strength used to drive the PWFA stage where the witness beam is generated. 
While both, electron driver beams produced by linacs as well as from LWFA, will exhibit some jitter in practice, LWFA-generated beams are especially prone to shot-to-shot fluctuations, often of the order of several percent in key properties. 
The most significant impact on the PWFA stage may arise from the drive beam charge density jitter, since drive beam density is directly responsible for excitation of the plasma wave via the Coulomb force.  
To explore this, we simulate a PWFA drive beam density jitter of $\pm 10\% $ around the baseline 
used previously. 
We keep the plasma photocathode $a_0$ constant at its baseline value, thus producing $Q_\mathrm{w} = 68$\,pC witness beams, propagate them through the transport line and examine the effect on the FEL performance.

\begin{figure*}[h!] 
    \centering
    \includegraphics[width = \textwidth]{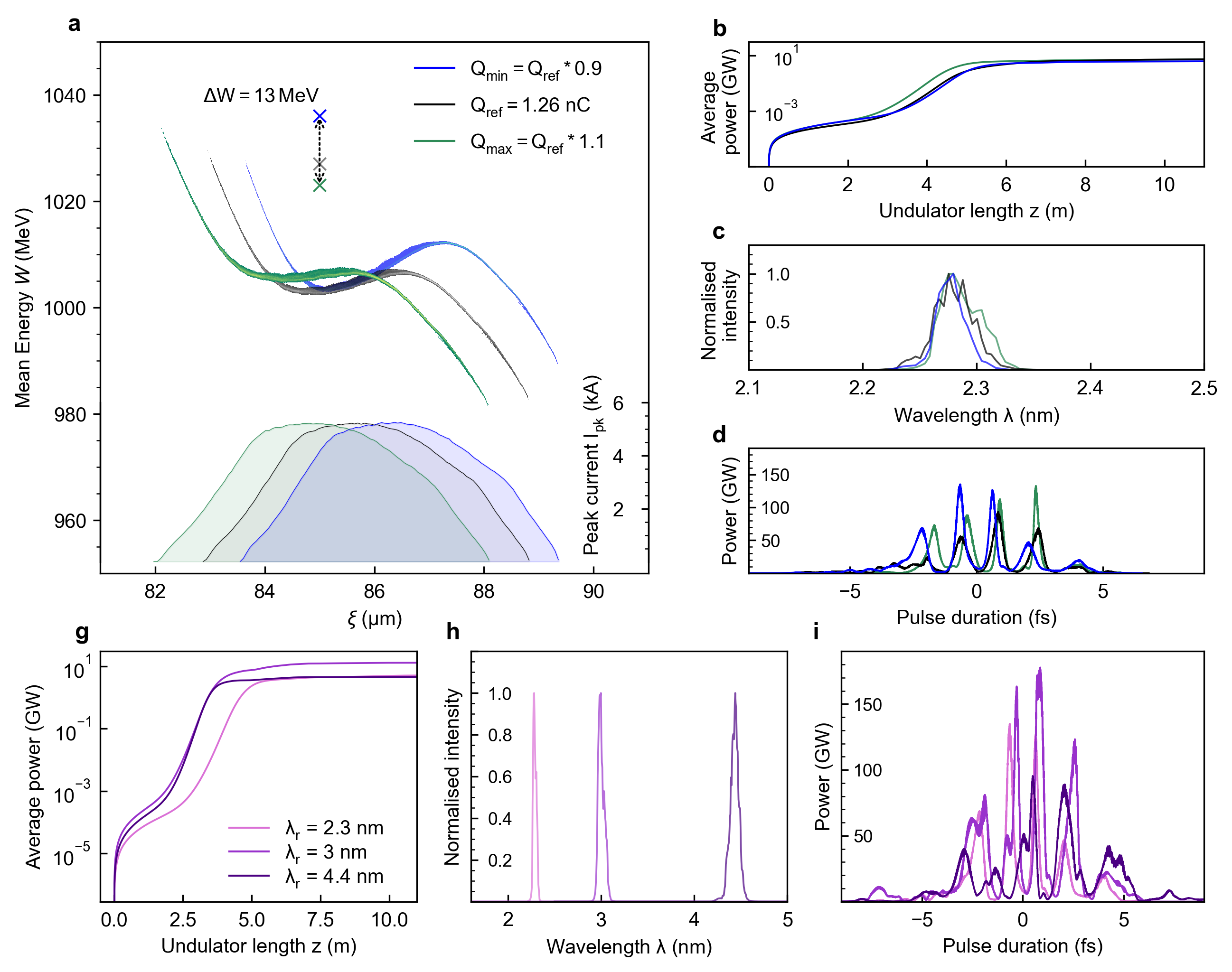}
    \caption{\textbf{Robustness and tunability across the water-window.} Impact of $\pm 10\,\%$ drive beam charge jitter on witness beam and FEL output, with a nominal drive beam charge $Q_{\mathrm{ref}} = 1.26$\,nC and witness charge $Q_{\mathrm{w}} = 68$\,pC. 
\textbf{a}, Witness beam longitudinal phase space at the plasma stage exit for three drive beam configurations: lower 
($Q_{\mathrm{min}} = Q_{\mathrm{ref}}\times 0.9$\, blue), nominal ($Q_{\mathrm{ref}}$, black), and high charge drive beam ($Q_{\mathrm{max}} = Q_{\mathrm{ref}}\times1.1$, green). 
Mean energy $W$ (solid lines) and peak current $I_{\mathrm{pk}}$ (shaded) are shown versus co-moving coordinate $\mathrm{\xi}$. Crosses indicate the mean energy of low charge ($\sim$3\,pC) beams trapped in the lower charge (blue), nominal (black) and higher charge (green) drive beam scenarios, with energy difference $\Delta W$ between them. 
\textbf{b}, Average FEL power gain as a function of undulator length $z$. 
\textbf{c}, Normalised spectral intensity around the fundamental FEL wavelength. 
\textbf{d}, Temporal pulse power profiles, showing robust pulse generation across all working points. 
\textbf{g–i}, Tuning of output wavelength by adjusting undulator parameter $K$ for the witness beam from the $Q_{\mathrm{ref}}$-case, shown for three target wavelengths: $\lambda_{\mathrm{r}} = 2.3$\,nm, 3\,nm, and 4.4\,nm. 
\textbf{g}, Average FEL power gain. 
\textbf{h}, Corresponding fundamental spectra. 
\textbf{i}, Temporal pulse structures for each wavelength configuration. 
}

    \label{fig5}
\end{figure*}

Fig. \ref{fig5}a shows that although the driver-induced variations in wakefield strength and size shift the trapping location and affect beam-loading, the effect on witness key properties is minimal. 
The peak current varies by $< 0.03$\,kA, slice emittance by $<2$\,nm\,rad and slice energy spread by $<0.001 \%$.
Moreover, the mean energy variation of the produced beams of $W \approx 1006 \pm 2$ MeV rms is surprisingly small. 
This becomes clear when compared to the energy variation of the non-beam-loaded ($\approx 3$ pC) beams released in the three different drive beam scenarios, indicated by the crosses in Fig. \ref{fig5}a. These exhibit a range of $W \approx 1029 \pm 7$ MeV.  
The remarkable, more than threefold reduction of energy variation when transitioning from unloaded to loaded operation is a 
further beneficial effect of beam-loading by the plasma photocathode. 
In the context of FEL application, the reduction in mean energy variation, together with the energy spread suppression, constitutes a mutually reinforcing, self-stabilising mechanism that improves robustness and quality of beam transport, as well as FEL operation. 
As a result, as shown in Fig. \ref{fig5}b-d, the XFEL performance after beam transport and focusing into the undulator remains strikingly robust in terms of power gain (Fig. \ref{fig5}b), intensity spectrum (Fig. \ref{fig5}c) and peak power at 100-GW-level and duration (Fig. \ref{fig5}d).

Such robustness in turn enables tunability of the XFEL radiation across the water window. 
Previous scenarios all aimed at a resonant wavelength of $\lambda_\mathrm{r} = 2.3$\,nm, i.e. the short-wavelength end of the water window, where achieving lasing is more challenging due to stringent beam quality requirements.   
Our hypothesis is therefore that if produced electron beams robustly lase at the more demanding wavelength across different $a_0$ and charge density 
working points, they should also support lasing at longer wavelengths. 
We explore this by tuning the resonant wavelength to $\lambda_\mathrm{r} = 2.3$\,nm, $\lambda_\mathrm{r} = 3$\,nm, and $\lambda_\mathrm{r} = 4.4$\,nm, respectively, by adjusting the undulator $B$-field amplitude 
(see Methods~\ref{sec:FELMethodes}). 
Figs. \ref{fig5}g-i show power gain, intensity spectra and pulse durations obtained for the three undulator settings and the reference electron beam. 
As expected, the power gain is even stronger, and saturation is reached even earlier when tuning the undulator to softer wavelengths. 
The intensity plots (here plotted in linear scale) show sharp profiles with $\Delta \lambda \approx 0.04\text{--}0.08$\,nm, and pulse durations are comparable in all cases, amounting to $\Delta \tau \approx 5$\,fs (FWHM) (see Extended Data Fig. 4 for pulse duration analysis). 
This shows that the XFEL can be tuned robustly across the whole water window.

\clearpage
\section{Discussion} \label{Discussion}
Our findings reveal a novel ultra-high-gain, high-efficiency regime of X-ray free-electron lasing, enabled by ultra-bright electron beams from a compact, plasma-based accelerator.  
First, the 
combined benefits of a plasma photocathode in dephasing-free plasma wakefield acceleration are exploited for beam generation and quality optimisation through direct beam-loading.
By raising the laser intensity, plasma photocathode-controlled charge release rates are enhanced, which increases the produced bunch current, introduces moderate emittance growth due to space charge, stabilises the mean energy, and decreases the energy chirp -- and hence the energy spread in the relevant cooperation length-scale slices due to beam-loading. 
Notably, the only beam parameter that worsens due to beam-loading is the emittance, while current, total beam energy, energy chirp and slice energy spread are all improved with regard to FEL performance. 
The 
excess emittance budget of the plasma photocathode allows such trade-offs to be made, whilst still keeping the slice emittance far below the lasing threshold. 

Then, the low projected energy spread, along with the novel self-stabilizing mechanism of the mean energy, facilitate a tunable beam transport line using standard beamline components, without compromising beam brightness.
This results in a scenario in which electron beams with unique properties can be focused into the undulator. 

Despite the initial emittance trade-off, the  averaged normalised slice emittance at the  $\varepsilon_\mathrm{n} < 50 $\,nm\,rad level, combined with multi-kA currents and sub-0.5\% relative projected energy spread, allows focusing to small beam radii with associated extremely high beam densities. 
The FEL is therefore quickly operating in an ultra-high-gain regime, effectively in overdrive.
The ultra-short gain lengths of few tens of cm allow growth of the photon field and FEL saturation to be reached robustly even in the absence of external focusing optics. 
Thus, the ultra-high brightness, focused beam density, and gain act in concert to achieve saturation within just a few metres of propagation in a standard undulator.
In addition, the very high FEL parameter yields very high extraction efficiency, and gigawatt-power, mJ-energy level pulses are produced.
Such ultra-high-gain enables robust operation far away from cliff-edge scenarios, and  generation of saturated FEL pulses despite potential shot-to-shot electron drive beam jitter.

Moreover, it opens up tunability of the FEL wavelength. 
Standard undulators are sufficient to reach the water-window, and undulator tunability can be used to  set the FEL wavelength working point across the whole water window. 
Additionally, fine-tuning of the resonant wavelength at the 0.1\,nm level can be achieved by controlling the electron witness beam energy via the plasma photocathode intensity. 
Such a functionality, achievable at contemporary seeded FELs by changing the seed laser wavelength \cite{allaria_tunability_2012},  could be useful for investigating atomic resonances or for extending strong-field quantum control \cite{Richter2024fermiQuantumControlNature} into the soft X-ray region. 

While large, conventional linac-based XFELs remain the gold standard for beam quality, compact laser-plasma-based accelerators have recently also begun to reach lasing thresholds at wavelengths ranging from several hundred down to a several tens of nanometres, albeit far away from reaching saturation powers.  
Our approach extends free-electron-lasing into new parameter regimes, and complements linac-based XFELs as well as other approaches via plasma-based accelerators. 
Compared to breakthroughs such as first plasma-accelerator based FEL proof-of-concepts \cite{wang_free-electron_nodate,pompili_free-electron_2022,LabatHZDRFEL2023,GallettiFELprospects2024} with thus far the shortest wavelength at 27\,nm and $\sim150$\,nJ pulse energies \cite{wang_free-electron_nodate}
,  our results show how an order of magnitude shorter wavelengths can be accessed, whilst reaching orders of magnitude stronger gain, reliable saturation, and a 3000-fold increase in pulse energy. 
This could be a key enabler for democratising FEL technology: if electron beams are reliably bright enough to achieve lasing by brute force rather than precision engineering, the demands on beam reproducibility can be relaxed, significantly lowering the technological precision capabilities required of setups and facilities.
In contrast to the 
escort beam dechirping scheme for plasma photocathodes -- primarily aimed at attosecond-\SI{}{\micro J}-class hard X-ray pulses \cite{habib_attosecond-angstrom_2023} -- our approach is explicitly designed for producing millijoule-class soft X-ray pulses.
It offers a more direct and technically simpler route, operates at significantly lower electron beam energies, and employs standard undulators, all within a compact setup -- while extending pulse energies into the millijoule regime.
Suitable drive beams can originate either from advanced linacs delivering currents >5\,kA, or from compact hybrid laser-plasma accelerators.
The latter has the potential to bridge the gap between compact but low-energy HHG sources and large-scale, high-energy linac-based soft XFELs targeting the water-window regime.
It is of significant interest for the core goal of the European Plasma Research Accelerator with Excellence in Applications (EuPRAXIA), one of the largest infrastructure projects on the European Strategy Forum on Research Infrastructures (ESFRI) roadmap \cite{assmann2020eupraxia}, which aims at a plasma-based FEL user facility with superior beam quality, but also for single research centres and even universities, thus democratising access to coherent soft XFELs.
parameter space will depend on the desired application.

\section{Methods}

\subsection{Plasma stage modelling}\label{sec:methods_plasma_stage}
The cylindrical-symmetric particle-in-cell (PIC) code FBPIC \cite{FBPIC} allows for high longitudinal and radial spatial resolution at a comparatively lower computational cost than fully explicit Cartesian 3D PIC codes. This makes it possible to employ a fully-resolved photocathode laser instead of using an envelope approximation, like in previous studies \cite{habib_plasma_2023}. Hence, this capability enabled high-fidelity modelling of centimetre-long PWFA stages across a wide parameter range. Furthermore, FBPIC uses the Pseudo-Spectral Analytical Time Domain (PSATD) algorithm which allows numerical Cherenkov (NC) radiation to be almost entirely mitigated when stencil order of 32 is used. Mitigation of NC radiation minimises artificial emittance growth in PIC codes \cite{PhysRevSTAB.16.021301-Lehe-2013} and is crucial when modelling electron beams with nm-rad normalised emittances.  Two spectral modes were used along with open box boundaries to capture relevant PWFA physics. The simulations were conducted in a co-moving frame which moves at the propagation speed of the drive beam ($\approx$ c).

A longitudinal resolution of $\mathrm{d}z$ = 66\,nm was used, corresponding to 12 cells per injection laser wavelength $\lambda_\mathrm{L} = 800$\,nm, and radial resolution of $\mathrm{d}r$ = \SI{0.4}{\micro\meter}. The simulation grid is comprised of $N$\textsubscript{z} $\times$ $N$\textsubscript{r} = 6629 $\times$ 469 = 3.1\,million cells. The electron drive beam was modelled with a Gaussian macroparticle distribution that is symmetric in the $z$- and $r$-axis with $\mathrm{\sigma_z}$ $ \times$ $\mathrm{\sigma_r}$ = \SI{25}{\um} $\times$ \SI{2.5}{\um}, normalised emittance $\varepsilon_{\mathrm{n,p}}=$ \SI{5}{\um} rad, energy spread $\sigma_{\mathrm{W,p}}=$ 5\,\% and energy of 1\,GeV. For smooth current profile 1\,million macroparticles were used for a driver charge of 1.26\,nC.

The drive beam propagates through the length of one simulation box before entering the plasma channel in the laboratory frame, which begins with a \SI{300}{\um} linear density upramp and is formed of $\mathrm{He}^+$ plasma electrons and ions modelled with 16 variable weight macroparticles per cell. This is followed by a \SI{85}{\milli\meter}-long uniform density profile, culminating in a \SI{5}{\milli\meter}-long $n_\mathrm{exit}\propto\cos^{2}$-shaped density downramp for electron beam extraction from the plasma stage. The plasma channel has a width $D \approx $ \SI{200}{\micro\meter} to support the unhindered formation and evolution of a blowout with radius $R\approx $ \SI{100}{\micro\meter} along the propagation direction -- a key condition for reliable and robust generation and acceleration of ultra-high brightness electron beams \cite{habib_plasma_2023}.  

The fully-resolved plasma photocathode injection laser has waist $w_0 =$ \SI{7}{um} and duration $\tau_0 =$ \SI{50}{fs} with Gaussian intensity profile in both planes with sub-mJ pulse energy. 
This allows 
one to tune the normalised laser vector potential $a_0 = e\lambda E_\mathrm{L}/(2\pi \mathrm{m_e} c^2)$ between the 0.0850--0.0960 range simply via the injector laser pulse energy, where $E_\mathrm{L}$ is the laser electric field. 
After \SI{2}{\milli\meter} of drive beam propagation in the laboratory frame, the laser is focused \SI{10}{\um} behind the centre of the blowout in the co-moving frame as defined by the longitudinal electric wakefield zero-crossing. When the laser ionises the $\mathrm{He}^+$ into $\mathrm{He}^{++}$, the resulting electrons forming the witness beam are also modelled with  16 variable weight macroparticles per cell. The field ionisation rates are based on the ADK model \cite{Ammosov_ADK_1986} and use the fully-resolved unaveraged laser field. 
The PWFA stage is sampled every \SI{100}{\um} in the laboratory frame to obtain field and particle data.

Witness beam properties in Fig. \ref{fig2} are calculated at the end of the downramp using only the core portion of the beam as defined by the FWHM of the longitudinal current distribution. This is justified because outside this section of the beam the low charge, chirped 'wings' do not contribute to lasing due to their lower current and high slice energy spread (see Supplementary Movie 1). 
Slice values are calculated using a slice width equal to the FEL cooperation length at a resonant wavelength of \SI{2.3}{nm} for the optimally loaded reference beam from Fig. \ref{fig1}f, which is $L_\mathrm{c}\approx$ \SI{115}{nm}. 
Whenever a single value is stated for a slice-dependent property, it has been computed as a current-weighted average across the individual slices.

\subsection{Beam transport modelling}\label{sec:TransportMethodes}

At the end of the downramp, the witness macroparticle distribution is prepared for the particle tracking code Elegant \cite{borland_elegant_2000}. 
Macroparticles in Elegant must have equal weight, but in FBPIC they have variable weight. 
Therefore we split each macroparticle from the FBPIC beam distribution into multiple macroparticles with equal weight and then resample 1 million of these new particles. 
We ensure that our resampling algorithm has a negligible effect on the beam properties. In Elegant, 3rd order matrices are used to model the beamline elements. 

The projected emittance and energy spread, current, $\alpha$ and $\beta$ in Fig. \ref{fig3} and Extended Fig. 2 are calculated internally in Elegant using the entire beam distribution including the chirped wings rather than just the beam core. 
Slice quantities were calculated using the slice analysis module within Elegant using 100 slices and were then weighted by beam slice current and averaged over the electron beam.

After initially drifting for 10\,cm, beams are captured by a set of permanent magnet quadrupoles (PMQs) which are necessary to collimate the witness beam upon exiting the plasma such that divergence is minimised. The beam then traverses a 1\,m drift before entering a set of electromagnetic quadrupoles (EMQs) for matching into the undulator. There is a final 0.5\,m of drift before the beam enters the undulator. 
The aim is to achieve a beta function of 1\,m at around 4\,m through the undulator, which is near the expected exponential gain region for most beams. For the `fixed beamline' runs the central energy is set to $W_0 = 1004$\,MeV for all beams to match the witness with the optimum brightness. 
For the `energy-adjusted' runs the central energy of the beamline was adjusted to the mean beam energy for each witness beam working point. This is done by setting the reference energy in Elegant, which adjusts the quadrupole field gradients $g$ according to quadrupole strength $k [\text{m}^{-2}] \approx 0.299 g[\text{T/m}]/\beta_{\mathrm{rel}} W \text{[GeV]}$, where $\beta_{\mathrm{rel}}$ is the relativistic beta-function associated with the beam velocity. 
The beamline was optimised using the build-in simplex algorithm for the configuration of the focusing (F) and defocusing (D) arrangement of the quadruples and configuration. During this optimisation it was found that using four PMQs increases the momentum acceptance range, so an F-D-F-D setup is used with strengths $k = 80.00, -79.91, 47.86, -9.02$ for the capturing line. 
The maximum quadrupole strength used ($k = 80$) corresponds to a field gradient of $g\approx$ 268\,T\,m$^{-1}$, which is well within reach of existing technology \cite{lim_adjustable_2005,LabatHZDRFEL2023}. 
Each quadrupole is \SI{10}{cm} in length and is separated by \SI{59}{mm}. After a \SI{1}{m} drift the beam enters the EMQ F-D-F triplet. The magnets are \SI{300}{mm} each in length and have strengths $k = 0.72, -1.73, 1.30$.  They are separated by \SI{10}{cm}. After a final \SI{0.5}{m} drift beams enter the undulator.

\subsection{FEL stage modelling}\label{sec:FELMethodes}

The unaveraged FEL code Puffin \cite{campbell_puffin_2012,Puffin-update-2018} was used for modelling the FEL stage, which is able to capture effects caused by current variations on the scale of the radiation wavelength such as Coherent Spontaneous Emission \cite{alotaibi_plasma_2020}. This requires sufficient macroparticles per resonant wavelength to avoid artificial current spikes on the scale of the cooperation length, which is more than are present in the beam distribution used in Elegant. 
Therefore the beam is first upsampled from 1 million to 10 million macroparticles via its joint cumulative distribution function. Realistic Poissonian shot noise is then added using the algorithm formulated in \cite{mcneil_unified_2003} to correctly simulate the startup radiation of the FEL.

A single $y$-polarised planar undulator is used with period $\lambda_\mathrm{u}=$ \SI{15}{mm} and peak field strength $B= 0.42$\,T, equivalent to an undulator parameter $K \approx 0.934 B[\mathrm{T}]\lambda_\mathrm{u}[\mathrm{cm}]= 0.59$.
No additional focusing is used beyond the natural undulator focusing that is present. The transverse resolution is \SI{5}{\micro\meter}, and the longitudinal resolution is 8 cells per resonant wavelength, which varies slightly due to the beam energy variation. This is sufficient to capture the fundamental, second and third FEL harmonics of the FEL radiation. 
The transverse simulation box size is $300\times300$ cells, which corresponds to 1.5\,mm and is sufficiently large to capture radiation diffraction across the wavelength range. 
The longitudinal box size is given in terms of the cooperation length, which again varies for each beam and is in the range of 100--170\,nm. 
The box length is set to 100 cooperation lengths and using 8 cells per resonant wavelength, leading to 30k$-$50k longitudinal cells. In the $\lambda_\mathrm{r} \approx 3$\,nm case in section \ref{robustness} the undulator parameters are $B = 0.74$\,T ($K = 1.04$) and in the $\lambda_\mathrm{r} \approx 4.4$\,nm case $B = 1.14 $\,T ($K = 1.59$), respectively.

The saturation length is calculated by finding the peak of the bunching parameter for the fundamental lasing harmonic. The gain length $L_\mathrm{g}$ refers to the FEL power gain length. Wavelength spectra in Figs. \ref{fig4}g and \ref{fig4}n are taken near the point of FEL saturation for each beam. 
For the calculation of $\rho_\mathrm{1D}$ in Figure \ref{fig4}, the definition given in \cite{ming_xie_design_1995} was used with the beam size $\sigma_\mathrm{x}$ taken from the $x$-plane. 
\\~\\
\noindent\textbf{Supplementary information.}
We provide a Supplementary Movie and Extended Data Figures to assist our central messages. 
\\~\\
\noindent\textbf{Acknowledgments.}
This research used resources of the National Energy Research
Scientific Computing Center, a DOE Office of Science User Facility
supported by the Office of Science of the U.S. Department of Energy
under Contract No. DE-AC02-05CH11231 using NERSC award
HEP-ERCAP0031797, and the project is supported by the European Research Council (ERC) under the EU Horizon 2020 research and innovation programme (NeXource: Next-generation Plasma-based Electron Beam Sources for High-brightness Photon Science, ERC Grant agreement No. 865877), and A.F.H. is supported by the STFC PWFA-FEL programme ST/S006214/1 and by STFC MoA 4132361. We thank Brian McNeil, David Dunning, Peter Williams, Edward Snedden, and Massimo Ferrario for useful discussions on FEL and EuPRAXIA. 
\\~\\
\noindent\textbf{Authors' contributions.}
A.F.H. and B.H. conceived the project and supervised its execution. L.B. carried through the start-to-end simulations under the supervision of A.F.H.; L.B. was leading data analysis with support from A.F.H.; L.B., A.F.H., T.H., T.W., D.C. and E.H. contributed to particle-in-cell simulations; L.B. and A.F.H. contributed to beam transport modelling; L.B. and A.F.H. contributed to FEL simulations. L.B., A.F.H. and B.H. wrote the manuscript; T.H. and T.W. provided useful comments to the manuscript draft. All authors contributed to discussing the results and finalising the manuscript.
\\~\\
\noindent\textbf{Conflict of interest.} 
A.F.H. and B.H. are inventors of a patent “Plasma Accelerator” WO2018069670A1 (status: published, Applicant: University of Strathclyde, with Radiabeam Technologies SME). The remaining authors declare no competing interests.
\\~\\
\noindent\textbf{Data availability.} 
The data that support the plots within this paper and other findings of this study are available from the corresponding author upon reasonable request.
\\~\\
\noindent\textbf{Code availability}

Analysis scripts and generated simulation data presented in this study are available from the corresponding authors upon reasonable request.

\renewcommand\refname{References}


\begin{thebibliography}{10}
\expandafter\ifx\csname url\endcsname\relax
  \def\url#1{\burl{#1}}\fi
\expandafter\ifx\csname urlprefix\endcsname\relax\def\urlprefix{URL }\fi
\providecommand{\bibinfo}[2]{#2}
\providecommand{\eprint}[2][]{\url{#2}}
\providecommand{\doi}[1]{\url{https://doi.org/#1}}

\bibitem{Pertot2017-HHG-at-water-window}
\bibinfo{author}{Pertot, Y.} \emph{et~al.}
\newblock \bibinfo{title}{Time-resolved x-ray absorption spectroscopy with a water window high-harmonic source}.
\newblock \emph{\bibinfo{journal}{Science}} \textbf{\bibinfo{volume}{355}}, \bibinfo{pages}{264--267} (\bibinfo{year}{2017}).
\newblock \urlprefix\url{https://www.science.org/doi/abs/10.1126/science.aah6114}.

\bibitem{WW_microscopy_kordel}
\bibinfo{author}{K\"{o}rdel, M.} \emph{et~al.}
\newblock \bibinfo{title}{Laboratory water-window x-ray microscopy}.
\newblock \emph{\bibinfo{journal}{Optica}} \textbf{\bibinfo{volume}{7}}, \bibinfo{pages}{658--674} (\bibinfo{year}{2020}).
\newblock \urlprefix\url{https://opg.optica.org/optica/abstract.cfm?URI=optica-7-6-658}.

\bibitem{alexander2024attosecond}
\bibinfo{author}{Alexander, O.} \emph{et~al.}
\newblock \bibinfo{title}{Attosecond impulsive stimulated x-ray raman scattering in liquid water}.
\newblock \emph{\bibinfo{journal}{Science Advances}} \textbf{\bibinfo{volume}{10}}, \bibinfo{pages}{eadp0841} (\bibinfo{year}{2024}).

\bibitem{li2024attosecond}
\bibinfo{author}{Li, S.} \emph{et~al.}
\newblock \bibinfo{title}{Attosecond-pump attosecond-probe x-ray spectroscopy of liquid water}.
\newblock \emph{\bibinfo{journal}{Science}} \textbf{\bibinfo{volume}{383}}, \bibinfo{pages}{1118--1122} (\bibinfo{year}{2024}).

\bibitem{guo2024experimental}
\bibinfo{author}{Guo, Z.} \emph{et~al.}
\newblock \bibinfo{title}{Experimental demonstration of attosecond pump--probe spectroscopy with an x-ray free-electron laser}.
\newblock \emph{\bibinfo{journal}{Nature Photonics}} \bibinfo{pages}{1--7} (\bibinfo{year}{2024}).

\bibitem{bharti_x-ray_2022}
\bibinfo{author}{Bharti, A.}, \bibinfo{author}{Turchet, A.} \& \bibinfo{author}{Marmiroli, B.}
\newblock \bibinfo{title}{X-{Ray} {Lithography} for {Nanofabrication}: {Is} {There} a {Future}?}
\newblock \emph{\bibinfo{journal}{Frontiers in Nanotechnology}} \textbf{\bibinfo{volume}{4}} (\bibinfo{year}{2022}).
\newblock \urlprefix\url{https://www.frontiersin.org/journals/nanotechnology/articles/10.3389/fnano.2022.835701}.

\bibitem{maldonado_x-ray_2016}
\bibinfo{author}{Maldonado, J.~R.} \& \bibinfo{author}{Peckerar, M.}
\newblock \bibinfo{title}{X-ray lithography: {Some} history, current status and future prospects}.
\newblock \emph{\bibinfo{journal}{Microelectronic Engineering}} \textbf{\bibinfo{volume}{161}}, \bibinfo{pages}{87--93} (\bibinfo{year}{2016}).
\newblock \urlprefix\url{https://www.sciencedirect.com/science/article/pii/S0167931716301757}.

\bibitem{krause1992high}
\bibinfo{author}{Krause, J.~L.}, \bibinfo{author}{Schafer, K.~J.} \& \bibinfo{author}{Kulander, K.~C.}
\newblock \bibinfo{title}{High-order harmonic generation from atoms and ions in the high intensity regime}.
\newblock \emph{\bibinfo{journal}{Physical Review Letters}} \textbf{\bibinfo{volume}{68}}, \bibinfo{pages}{3535} (\bibinfo{year}{1992}).

\bibitem{l1993high}
\bibinfo{author}{L’Huillier, A.} \& \bibinfo{author}{Balcou, P.}
\newblock \bibinfo{title}{High-order harmonic generation in rare gases with a 1-ps 1053-nm laser}.
\newblock \emph{\bibinfo{journal}{Physical Review Letters}} \textbf{\bibinfo{volume}{70}}, \bibinfo{pages}{774} (\bibinfo{year}{1993}).

\bibitem{paul2001observation}
\bibinfo{author}{Paul, P.-M.} \emph{et~al.}
\newblock \bibinfo{title}{Observation of a train of attosecond pulses from high harmonic generation}.
\newblock \emph{\bibinfo{journal}{Science}} \textbf{\bibinfo{volume}{292}}, \bibinfo{pages}{1689--1692} (\bibinfo{year}{2001}).

\bibitem{FLASH_ackermann_2007}
\bibinfo{author}{Ackermann, W.} \emph{et~al.}
\newblock \bibinfo{title}{Operation of a free-electron laser from the extreme ultraviolet to the water window}.
\newblock \emph{\bibinfo{journal}{Nature Photonics}} \textbf{\bibinfo{volume}{1}}, \bibinfo{pages}{336--342} (\bibinfo{year}{2007}).
\newblock \urlprefix\url{https://www.nature.com/articles/nphoton.2007.76}.
\newblock \bibinfo{note}{Publisher: Nature Publishing Group}.

\bibitem{first-lasing-lcls-2010}
\bibinfo{author}{Emma, P.} \emph{et~al.}
\newblock \bibinfo{title}{First lasing and operation of an {\aa}ngstrom-wavelength free-electron laser}.
\newblock \emph{\bibinfo{journal}{Nature photonics}} \textbf{\bibinfo{volume}{4}}, \bibinfo{pages}{641--647} (\bibinfo{year}{2010}).
\newblock \urlprefix\url{http://www.nature.com/nphoton/journal/v4/n9/abs/nphoton.2010.176.html}.

\bibitem{HHG-conversion-Shiner-2009}
\bibinfo{author}{Shiner, A.~D.} \emph{et~al.}
\newblock \bibinfo{title}{Wavelength scaling of high harmonic generation efficiency}.
\newblock \emph{\bibinfo{journal}{Phys. Rev. Lett.}} \textbf{\bibinfo{volume}{103}}, \bibinfo{pages}{073902} (\bibinfo{year}{2009}).
\newblock \urlprefix\url{https://link.aps.org/doi/10.1103/PhysRevLett.103.073902}.

\bibitem{li_attosecond_2020}
\bibinfo{author}{Li, J.} \emph{et~al.}
\newblock \bibinfo{title}{Attosecond science based on high harmonic generation from gases and solids}.
\newblock \emph{\bibinfo{journal}{Nature Communications}} \textbf{\bibinfo{volume}{11}}, \bibinfo{pages}{2748} (\bibinfo{year}{2020}).
\newblock \urlprefix\url{https://www.nature.com/articles/s41467-020-16480-6}.
\newblock \bibinfo{note}{Publisher: Nature Publishing Group}.

\bibitem{huang_review_2007}
\bibinfo{author}{Huang, Z.} \& \bibinfo{author}{Kim, K.-J.}
\newblock \bibinfo{title}{Review of x-ray free-electron laser theory}.
\newblock \emph{\bibinfo{journal}{Physical Review Special Topics - Accelerators and Beams}} \textbf{\bibinfo{volume}{10}}, \bibinfo{pages}{034801} (\bibinfo{year}{2007}).
\newblock \urlprefix\url{https://link.aps.org/doi/10.1103/PhysRevSTAB.10.034801}.

\bibitem{BrianNatPhoton2010}
\bibinfo{author}{McNeil, B. W.~J.}
\newblock \bibinfo{title}{X-ray free-electron lasers}.
\newblock \emph{\bibinfo{journal}{Nature Photonics}} \textbf{\bibinfo{volume}{4}}, \bibinfo{pages}{814--821} (\bibinfo{year}{2010}).
\newblock \urlprefix\url{http://www.nature.com/nphoton/journal/v4/n12/abs/nphoton.2010.239.html}.

\bibitem{DiMitri:2014wpa}
\bibinfo{author}{Di~Mitri, S.} \& \bibinfo{author}{Cornacchia, M.}
\newblock \bibinfo{title}{{Electron beam brightness in linac drivers for free-electron-lasers}}.
\newblock \emph{\bibinfo{journal}{Phys. Rept.}} \textbf{\bibinfo{volume}{539}}, \bibinfo{pages}{1--48} (\bibinfo{year}{2014}).
\newblock \urlprefix\url{doi:10.1016/j.physrep.2014.01.005}.

\bibitem{rosenzweig2024high}
\bibinfo{author}{Rosenzweig, J.~B.} \emph{et~al.}
\newblock \bibinfo{title}{A high-flux compact x-ray free-electron laser for next-generation chip metrology needs}.
\newblock \emph{\bibinfo{journal}{Instruments}} \textbf{\bibinfo{volume}{8}}, \bibinfo{pages}{19} (\bibinfo{year}{2024}).

\bibitem{Schlenvoigt2008}
\bibinfo{author}{Schlenvoigt, H.-P.} \emph{et~al.}
\newblock \bibinfo{title}{A compact synchrotron radiation source driven by a laser-plasma wakefield accelerator}.
\newblock \emph{\bibinfo{journal}{Nature physics}} \textbf{\bibinfo{volume}{4}}, \bibinfo{pages}{130--133} (\bibinfo{year}{2008}).
\newblock \urlprefix\url{http://www.nature.com/nphys/journal/v4/n2/pdf/nphys811.pdf}.

\bibitem{Fuchs2009}
\bibinfo{author}{Fuchs, M.} \emph{et~al.}
\newblock \bibinfo{title}{Laser-driven soft-x-ray undulator source}.
\newblock \emph{\bibinfo{journal}{Nature physics}} \textbf{\bibinfo{volume}{5}}, \bibinfo{pages}{826--829} (\bibinfo{year}{2009}).

\bibitem{maier2020water}
\bibinfo{author}{Maier, A.~R.} \emph{et~al.}
\newblock \bibinfo{title}{Water-window x-ray pulses from a laser-plasma driven undulator}.
\newblock \emph{\bibinfo{journal}{Scientific reports}} \textbf{\bibinfo{volume}{10}}, \bibinfo{pages}{5634} (\bibinfo{year}{2020}).

\bibitem{wang_free-electron_nodate}
\bibinfo{author}{Wang, W.} \emph{et~al.}
\newblock \bibinfo{title}{Free-electron lasing at 27 nanometres based on a laser wakefield accelerator}.
\newblock \emph{\bibinfo{journal}{Nature}} \textbf{\bibinfo{volume}{595}}, \bibinfo{pages}{516--520} (\bibinfo{year}{2021}).
\newblock \urlprefix\url{doi:10.1038/s41586-021-03678-x}.

\bibitem{LabatHZDRFEL2023}
\bibinfo{author}{Labat, M.} \emph{et~al.}
\newblock \bibinfo{title}{Seeded free-electron laser driven by a compact laser plasma accelerator}.
\newblock \emph{\bibinfo{journal}{Nature Photonics}} \textbf{\bibinfo{volume}{17}}, \bibinfo{pages}{150--156} (\bibinfo{year}{2023}).
\newblock \urlprefix\url{https://doi.org/10.1038/s41566-022-01104-w}.

\bibitem{pompili_free-electron_2022}
\bibinfo{author}{Pompili, R.} \emph{et~al.}
\newblock \bibinfo{title}{Free-electron lasing with compact beam-driven plasma wakefield accelerator}.
\newblock \emph{\bibinfo{journal}{Nature}} \textbf{\bibinfo{volume}{605}}, \bibinfo{pages}{659--662} (\bibinfo{year}{2022}).
\newblock \urlprefix\url{doi:10.1038/s41586-022-04589-1}.

\bibitem{assmann2020eupraxia}
\bibinfo{author}{Assmann, R.} \emph{et~al.}
\newblock \bibinfo{title}{Eupraxia conceptual design report}.
\newblock \emph{\bibinfo{journal}{The European Physical Journal Special Topics}} \textbf{\bibinfo{volume}{229}}, \bibinfo{pages}{3675--4284} (\bibinfo{year}{2020}).

\bibitem{emma2021free}
\bibinfo{author}{Emma, C.} \emph{et~al.}
\newblock \bibinfo{title}{Free electron lasers driven by plasma accelerators: status and near-term prospects}.
\newblock \emph{\bibinfo{journal}{High Power Laser Science and Engineering}} \textbf{\bibinfo{volume}{9}}, \bibinfo{pages}{e57} (\bibinfo{year}{2021}).

\bibitem{galletti2024prospects}
\bibinfo{author}{Galletti, M.} \emph{et~al.}
\newblock \bibinfo{title}{Prospects for free-electron lasers powered by plasma-wakefield-accelerated beams}.
\newblock \emph{\bibinfo{journal}{Nature Photonics}} \textbf{\bibinfo{volume}{18}}, \bibinfo{pages}{780--791} (\bibinfo{year}{2024}).

\bibitem{lindstrom2025beam}
\bibinfo{author}{Lindstr{\o}m, C.} \emph{et~al.}
\newblock \bibinfo{title}{Beam-driven plasma-wakefield acceleration}.
\newblock \emph{\bibinfo{journal}{arXiv preprint arXiv:2504.05558}}  (\bibinfo{year}{2025}).

\bibitem{hidding_ultracold_2012}
\bibinfo{author}{Hidding, B.} \emph{et~al.}
\newblock \bibinfo{title}{Ultracold electron bunch generation via plasma photocathode emission and acceleration in a beam-driven plasma blowout}.
\newblock \emph{\bibinfo{journal}{Phys. Rev. Lett.}} \textbf{\bibinfo{volume}{108}}, \bibinfo{pages}{035001} (\bibinfo{year}{2012}).
\newblock \urlprefix\url{doi:10.1103/PhysRevLett.108.035001}.

\bibitem{manahan_single-stage_2017}
\bibinfo{author}{{G.G. Manahan and A.F. Habib,}} \emph{et~al.}
\newblock \bibinfo{title}{Single-stage plasma-based correlated energy spread compensation for ultrahigh {6D} brightness electron beams}.
\newblock \emph{\bibinfo{journal}{Nat. Commun.}} \textbf{\bibinfo{volume}{8}} (\bibinfo{year}{2017}).
\newblock \urlprefix\url{doi:10.1038/ncomms15705}.

\bibitem{habib_attosecond-angstrom_2023}
\bibinfo{author}{Habib, A.~F.} \emph{et~al.}
\newblock \bibinfo{title}{Attosecond-{Angstrom} free-electron-laser towards the cold beam limit}.
\newblock \emph{\bibinfo{journal}{Nature Communications}} \textbf{\bibinfo{volume}{14}}, \bibinfo{pages}{1054} (\bibinfo{year}{2023}).
\newblock \urlprefix\url{https://www.nature.com/articles/s41467-023-36592-z}.

\bibitem{chen_acceleration_1985}
\bibinfo{author}{Chen, P.}, \bibinfo{author}{Dawson, J.~M.}, \bibinfo{author}{Huff, R.~W.} \& \bibinfo{author}{Katsouleas, T.}
\newblock \bibinfo{title}{Acceleration of electrons by the interaction of a bunched electron beam with a plasma}.
\newblock \emph{\bibinfo{journal}{Phys. Rev. Lett.}} \textbf{\bibinfo{volume}{54}}, \bibinfo{pages}{693--696} (\bibinfo{year}{1985}).
\newblock \urlprefix\url{doi:10.1103/PhysRevLett.54.693}.

\bibitem{rosenzweig_nonlinear_1987}
\bibinfo{author}{Rosenzweig, J.~B.}
\newblock \bibinfo{title}{Nonlinear plasma dynamics in the plasma wake-field accelerator}.
\newblock \emph{\bibinfo{journal}{Physical Review Letters}} \textbf{\bibinfo{volume}{58}}, \bibinfo{pages}{555--558} (\bibinfo{year}{1987}).
\newblock \urlprefix\url{https://link.aps.org/doi/10.1103/PhysRevLett.58.555}.
\newblock \bibinfo{note}{Publisher: American Physical Society}.

\bibitem{rosenzweig_experimental_1988}
\bibinfo{author}{Rosenzweig, J.~B.} \emph{et~al.}
\newblock \bibinfo{title}{Experimental {Observation} of {Plasma} {Wake}-{Field} {Acceleration}}.
\newblock \emph{\bibinfo{journal}{Physical Review Letters}} \textbf{\bibinfo{volume}{61}}, \bibinfo{pages}{98--101} (\bibinfo{year}{1988}).
\newblock \urlprefix\url{https://link.aps.org/doi/10.1103/PhysRevLett.61.98}.
\newblock \bibinfo{note}{Publisher: American Physical Society}.

\bibitem{blumenfeld_energy_2007}
\bibinfo{author}{Blumenfeld, I.} \emph{et~al.}
\newblock \bibinfo{title}{Energy doubling of 42 {GeV} electrons in a metre-scale plasma wakefield accelerator}.
\newblock \emph{\bibinfo{journal}{Nature}} \textbf{\bibinfo{volume}{445}}, \bibinfo{pages}{741--744} (\bibinfo{year}{2007}).
\newblock \urlprefix\url{doi:10.1038/nature05538}.

\bibitem{Litos2014}
\bibinfo{author}{Litos, M.} \emph{et~al.}
\newblock \bibinfo{title}{High-efficiency acceleration of an electron beam in a plasma wakefield accelerator}.
\newblock \emph{\bibinfo{journal}{Nature}} \textbf{\bibinfo{volume}{515}}, \bibinfo{pages}{92--95} (\bibinfo{year}{2014}).
\newblock \urlprefix\url{http://dx.doi.org/10.1038/nature13882}.

\bibitem{Katsouleas:1987yd}
\bibinfo{author}{Katsouleas, T.~C.}, \bibinfo{author}{Wilks, S.}, \bibinfo{author}{Chen, P.}, \bibinfo{author}{Dawson, J.~M.} \& \bibinfo{author}{Su, J.~J.}
\newblock \bibinfo{title}{Beam loading in plasma accelerators}.
\newblock \emph{\bibinfo{journal}{Part. Accel.}} \textbf{\bibinfo{volume}{22}}, \bibinfo{pages}{81--99} (\bibinfo{year}{1987}).

\bibitem{tzoufras_beam_2008}
\bibinfo{author}{Tzoufras, M.} \emph{et~al.}
\newblock \bibinfo{title}{Beam {Loading} in the {Nonlinear} {Regime} of {Plasma}-{Based} {Acceleration}}.
\newblock \emph{\bibinfo{journal}{Physical Review Letters}} \textbf{\bibinfo{volume}{101}}, \bibinfo{pages}{145002} (\bibinfo{year}{2008}).
\newblock \urlprefix\url{https://link.aps.org/doi/10.1103/PhysRevLett.101.145002}.
\newblock \bibinfo{note}{Publisher: American Physical Society}.

\bibitem{habib_plasma_2023}
\bibinfo{author}{Habib, A.~F.} \emph{et~al.}
\newblock \bibinfo{title}{Plasma {Photocathodes}}.
\newblock \emph{\bibinfo{journal}{Annalen der Physik}} \textbf{\bibinfo{volume}{535}}, \bibinfo{pages}{2200655} (\bibinfo{year}{2023}).
\newblock \urlprefix\url{https://onlinelibrary.wiley.com/doi/10.1002/andp.202200655}.

\bibitem{deng_generation_2019}
\bibinfo{author}{Deng, A.} \emph{et~al.}
\newblock \bibinfo{title}{Generation and acceleration of electron bunches from a plasma photocathode}.
\newblock \emph{\bibinfo{journal}{Nat. Phys.}} \textbf{\bibinfo{volume}{15}}, \bibinfo{pages}{1--5} (\bibinfo{year}{2019}).
\newblock \urlprefix\url{doi:10.1038/s41567-019-0610-9}.

\bibitem{uferNutterToBePublished}
\bibinfo{author}{Ufer, P.}, \bibinfo{author}{Nutter, A.} \emph{et~al.}
\newblock \bibinfo{title}{To be published}.

\bibitem{HiddingPRL2010PhysRevLett.104.195002short}
\bibinfo{author}{Hidding, B.} \emph{et~al.}
\newblock \bibinfo{title}{Monoenergetic energy doubling in a hybrid laser-plasma wakefield accelerator}.
\newblock \emph{\bibinfo{journal}{Phys. Rev. Lett.}} \textbf{\bibinfo{volume}{104}}, \bibinfo{pages}{195002} (\bibinfo{year}{2010}).

\bibitem{kurz_demonstration_2021}
\bibinfo{author}{Kurz, T.} \emph{et~al.}
\newblock \bibinfo{title}{Demonstration of a compact plasma accelerator powered by laser-accelerated electron beams}.
\newblock \emph{\bibinfo{journal}{Nature Communications}} \textbf{\bibinfo{volume}{12}}, \bibinfo{pages}{2895} (\bibinfo{year}{2021}).
\newblock \urlprefix\url{http://www.nature.com/articles/s41467-021-23000-7}.

\bibitem{FoersterPRX2022}
\bibinfo{author}{Foerster, F.~M.} \emph{et~al.}
\newblock \bibinfo{title}{Stable and high-quality electron beams from staged laser and plasma wakefield accelerators}.
\newblock \emph{\bibinfo{journal}{Phys. Rev. X}} \textbf{\bibinfo{volume}{12}}, \bibinfo{pages}{041016} (\bibinfo{year}{2022}).
\newblock \urlprefix\url{https://link.aps.org/doi/10.1103/PhysRevX.12.041016}.

\bibitem{FBPIC}
\bibinfo{author}{Lehe, R.}, \bibinfo{author}{Kirchen, M.}, \bibinfo{author}{Andriyash, I.~A.}, \bibinfo{author}{Godfrey, B.~B.} \& \bibinfo{author}{Vay, J.-L.}
\newblock \bibinfo{title}{A spectral, quasi-cylindrical and dispersion-free {Particle}-{In}-{Cell} algorithm}.
\newblock \emph{\bibinfo{journal}{Computer Physics Communications}} \textbf{\bibinfo{volume}{203}}, \bibinfo{pages}{66--82} (\bibinfo{year}{2016}).
\newblock \urlprefix\url{https://linkinghub.elsevier.com/retrieve/pii/S0010465516300224}.

\bibitem{ThermalEmittancePhysRevSTAB.17.101301Schroeder2014}
\bibinfo{author}{Schroeder, C.~B.} \emph{et~al.}
\newblock \bibinfo{title}{Thermal emittance from ionization-induced trapping in plasma accelerators}.
\newblock \emph{\bibinfo{journal}{Phys. Rev. ST Accel. Beams}} \textbf{\bibinfo{volume}{17}}, \bibinfo{pages}{101301} (\bibinfo{year}{2014}).
\newblock \urlprefix\url{doi:10.1103/PhysRevSTAB.17.101301}.

\bibitem{Floettmann2003}
\bibinfo{author}{Floettmann, K.}
\newblock \bibinfo{title}{Some basic features of the beam emittance}.
\newblock \emph{\bibinfo{journal}{Phys. Rev. ST Accel. Beams}} \textbf{\bibinfo{volume}{6}}, \bibinfo{pages}{034202} (\bibinfo{year}{2003}).
\newblock \urlprefix\url{https://link.aps.org/doi/10.1103/PhysRevSTAB.6.034202}.

\bibitem{Mehrling2012}
\bibinfo{author}{Mehrling, T.}, \bibinfo{author}{Grebenyuk, J.}, \bibinfo{author}{Tsung, F.~S.}, \bibinfo{author}{Floettmann, K.} \& \bibinfo{author}{Osterhoff, J.}
\newblock \bibinfo{title}{Transverse emittance growth in staged laser-wakefield acceleration}.
\newblock \emph{\bibinfo{journal}{Phys. Rev. ST Accel. Beams}} \textbf{\bibinfo{volume}{15}}, \bibinfo{pages}{111303--} (\bibinfo{year}{2012}).
\newblock \urlprefix\url{http://link.aps.org/doi/10.1103/PhysRevSTAB.15.111303}.

\bibitem{migliorati_intrinsic_2013}
\bibinfo{author}{Migliorati, M.} \emph{et~al.}
\newblock \bibinfo{title}{Intrinsic normalized emittance growth in laser-driven electron accelerators}.
\newblock \emph{\bibinfo{journal}{Physical Review Special Topics - Accelerators and Beams}} \textbf{\bibinfo{volume}{16}}, \bibinfo{pages}{011302} (\bibinfo{year}{2013}).
\newblock \urlprefix\url{https://link.aps.org/doi/10.1103/PhysRevSTAB.16.011302}.

\bibitem{andre2018control}
\bibinfo{author}{Andr{\'e}, T.} \emph{et~al.}
\newblock \bibinfo{title}{Control of laser plasma accelerated electrons for light sources}.
\newblock \emph{\bibinfo{journal}{Nature communications}} \textbf{\bibinfo{volume}{9}}, \bibinfo{pages}{1334} (\bibinfo{year}{2018}).
\newblock \urlprefix\url{https://doi.org/10.1038/s41467-018-03776-x}.

\bibitem{PhysRevX.10.031039}
\bibinfo{author}{Maier, A.~R.} \emph{et~al.}
\newblock \bibinfo{title}{Decoding sources of energy variability in a laser-plasma accelerator}.
\newblock \emph{\bibinfo{journal}{Phys. Rev. X}} \textbf{\bibinfo{volume}{10}}, \bibinfo{pages}{031039} (\bibinfo{year}{2020}).
\newblock \urlprefix\url{https://link.aps.org/doi/10.1103/PhysRevX.10.031039}.

\bibitem{Carl2016achromat}
\bibinfo{author}{Lindstr\o{}m, C.~A.} \& \bibinfo{author}{Adli, E.}
\newblock \bibinfo{title}{Design of general apochromatic drift-quadrupole beam lines}.
\newblock \emph{\bibinfo{journal}{Phys. Rev. Accel. Beams}} \textbf{\bibinfo{volume}{19}}, \bibinfo{pages}{071002} (\bibinfo{year}{2016}).
\newblock \urlprefix\url{https://link.aps.org/doi/10.1103/PhysRevAccelBeams.19.071002}.

\bibitem{borland_elegant_2000}
\bibinfo{author}{Borland, M.}
\newblock \emph{\bibinfo{title}{elegant: {A} {Flexible} {SDDS}-{Compliant} {Code} for {Accelerator} {Simulation}}} (\bibinfo{year}{2000}).

\bibitem{campbell_puffin_2012}
\bibinfo{author}{Campbell, L.~T.} \& \bibinfo{author}{McNeil, B. W.~J.}
\newblock \bibinfo{title}{Puffin: {A} three dimensional, unaveraged free electron laser simulation code}.
\newblock \emph{\bibinfo{journal}{Physics of Plasmas}} \textbf{\bibinfo{volume}{19}}, \bibinfo{pages}{093119} (\bibinfo{year}{2012}).
\newblock \urlprefix\url{http://aip.scitation.org/doi/10.1063/1.4752743}.

\bibitem{Puffin-update-2018}
\bibinfo{author}{Campbell, L.}, \bibinfo{author}{McNeil, B.}, \bibinfo{author}{Smith, J.} \& \bibinfo{author}{Traczykowski, P.}
\newblock \bibinfo{editor}{Koscielniak, S.}, \bibinfo{editor}{Satogata, T.}, \bibinfo{editor}{Schaa, V.} \& \bibinfo{editor}{Thomson, J.} (eds) \emph{\bibinfo{title}{An updated description of the fel simulation code puffin}}.
\newblock (eds \bibinfo{editor}{Koscielniak, S.}, \bibinfo{editor}{Satogata, T.}, \bibinfo{editor}{Schaa, V.} \& \bibinfo{editor}{Thomson, J.}) \emph{\bibinfo{booktitle}{Proceedings, 9th International Particle Accelerator Conference (IPAC 2018)}} (\bibinfo{year}{2018}).
\newblock \bibinfo{note}{9th International Particle Accelerator Conference, IPAC 2018 ; Conference date: 29-04-2018 Through 04-05-2018}.

\bibitem{ming_xie_design_1995}
\bibinfo{author}{{M. Xie}}
\newblock \emph{\bibinfo{title}{Design optimization for an {X}-ray free electron laser driven by {SLAC} linac}}, Vol.~\bibinfo{volume}{1}, \bibinfo{pages}{183--185} (\bibinfo{publisher}{IEEE}, \bibinfo{address}{Dallas, TX, USA}, \bibinfo{year}{1995}).
\newblock \urlprefix\url{doi:10.1109/PAC.1995.504603}.

\bibitem{PhysRevA.40.4467-Bonifacio-Superrad}
\bibinfo{author}{Bonifacio, R.}, \bibinfo{author}{McNeil, B. W.~J.} \& \bibinfo{author}{Pierini, P.}
\newblock \bibinfo{title}{Superradiance in the high-gain free-electron laser}.
\newblock \emph{\bibinfo{journal}{Phys. Rev. A}} \textbf{\bibinfo{volume}{40}}, \bibinfo{pages}{4467--4475} (\bibinfo{year}{1989}).
\newblock \urlprefix\url{https://link.aps.org/doi/10.1103/PhysRevA.40.4467}.

\bibitem{allaria_tunability_2012}
\bibinfo{author}{Allaria, E.} \emph{et~al.}
\newblock \bibinfo{title}{Tunability experiments at the {FERMI}@{Elettra} free-electron laser}.
\newblock \emph{\bibinfo{journal}{New Journal of Physics}} \textbf{\bibinfo{volume}{14}}, \bibinfo{pages}{113009} (\bibinfo{year}{2012}).
\newblock \urlprefix\url{https://dx.doi.org/10.1088/1367-2630/14/11/113009}.
\newblock \bibinfo{note}{Publisher: IOP Publishing}.

\bibitem{Richter2024fermiQuantumControlNature}
\bibinfo{author}{Richter, F.} \emph{et~al.}
\newblock \bibinfo{title}{Strong-field quantum control in the extreme ultraviolet domain using pulse shaping}.
\newblock \emph{\bibinfo{journal}{Nature}} \textbf{\bibinfo{volume}{636}}, \bibinfo{pages}{337--341} (\bibinfo{year}{2024}).
\newblock \urlprefix\url{https://doi.org/10.1038/s41586-024-08209-y}.

\bibitem{GallettiFELprospects2024}
\bibinfo{author}{Galletti, M.} \emph{et~al.}
\newblock \bibinfo{title}{Prospects for free-electron lasers powered by plasma-wakefield-accelerated beams}.
\newblock \emph{\bibinfo{journal}{Nature Photonics}} \textbf{\bibinfo{volume}{18}}, \bibinfo{pages}{780--791} (\bibinfo{year}{2024}).
\newblock \urlprefix\url{https://doi.org/10.1038/s41566-024-01474-3}.

\bibitem{PhysRevSTAB.16.021301-Lehe-2013}
\bibinfo{author}{Lehe, R.}, \bibinfo{author}{Lifschitz, A.}, \bibinfo{author}{Thaury, C.}, \bibinfo{author}{Malka, V.} \& \bibinfo{author}{Davoine, X.}
\newblock \bibinfo{title}{Numerical growth of emittance in simulations of laser-wakefield acceleration}.
\newblock \emph{\bibinfo{journal}{Phys. Rev. ST Accel. Beams}} \textbf{\bibinfo{volume}{16}}, \bibinfo{pages}{021301} (\bibinfo{year}{2013}).
\newblock \urlprefix\url{https://link.aps.org/doi/10.1103/PhysRevSTAB.16.021301}.

\bibitem{Ammosov_ADK_1986}
\bibinfo{author}{Ammosov, M.~V.}, \bibinfo{author}{Delone, N.~B.} \& \bibinfo{author}{Krainov, V.~P.}
\newblock \bibinfo{title}{Tunnel ionization of complex atoms and atomic ions in electromagnetic field}.
\newblock \emph{\bibinfo{journal}{Proc. SPIE.}} \textbf{\bibinfo{volume}{0664}} (\bibinfo{year}{1986}).

\bibitem{lim_adjustable_2005}
\bibinfo{author}{Lim, J.~K.} \emph{et~al.}
\newblock \bibinfo{title}{Adjustable, short focal length permanent-magnet quadrupole based electron beam final focus system}.
\newblock \emph{\bibinfo{journal}{Physical Review Special Topics - Accelerators and Beams}} \textbf{\bibinfo{volume}{8}}, \bibinfo{pages}{072401} (\bibinfo{year}{2005}).
\newblock \urlprefix\url{https://link.aps.org/doi/10.1103/PhysRevSTAB.8.072401}.
\newblock \bibinfo{note}{Publisher: American Physical Society}.

\bibitem{alotaibi_plasma_2020}
\bibinfo{author}{Alotaibi, B.~M.} \emph{et~al.}
\newblock \bibinfo{title}{Plasma wakefield accelerator driven coherent spontaneous emission from an energy chirped electron pulse}.
\newblock \emph{\bibinfo{journal}{New Journal of Physics}} \textbf{\bibinfo{volume}{22}}, \bibinfo{pages}{013037} (\bibinfo{year}{2020}).
\newblock \urlprefix\url{https://iopscience.iop.org/article/10.1088/1367-2630/ab64ae}.

\bibitem{mcneil_unified_2003}
\bibinfo{author}{McNeil, B. W.~J.}, \bibinfo{author}{Poole, M.~W.} \& \bibinfo{author}{Robb, G. R.~M.}
\newblock \bibinfo{title}{Unified model of electron beam shot noise and coherent spontaneous emission in the helical wiggler free electron laser}.
\newblock \emph{\bibinfo{journal}{Physical Review Special Topics - Accelerators and Beams}} \textbf{\bibinfo{volume}{6}}, \bibinfo{pages}{070701} (\bibinfo{year}{2003}).
\newblock \urlprefix\url{https://link.aps.org/doi/10.1103/PhysRevSTAB.6.070701}.

\end{thebibliography}

\section{Extended Data}

\begin{sidewaysfigure}[h]
  \centering
  \includegraphics[width=\textheight]{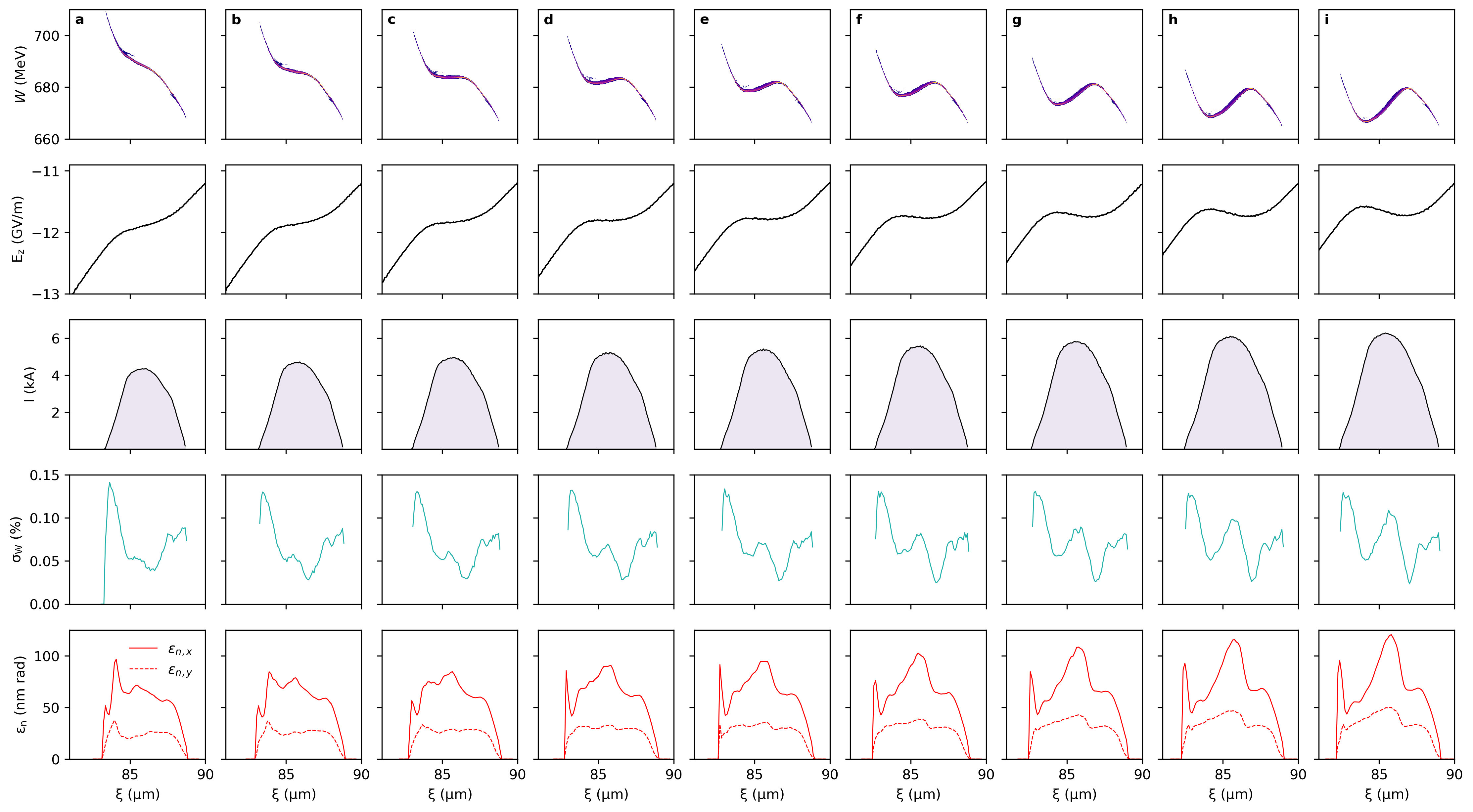}
  \caption*{\textbf{Extended Data Fig. 1} From top to bottom row: Longitudinal phase space, longitudinal electric field $E_z$, current $I$, slice energy spread $\sigma_\mathrm{W,s}$ and slice normalised emittance $\varepsilon_\mathrm{n,s}$ for each beam plotted in Fig. \ref{fig2}. Beam charge, current, emittance and energy spread increase from \textbf{a-i} and beams pass from an underloaded to an overloaded regime. The reference beam with optimum loading is in column d with 68\,pC.}
  \label{ext_fig1}
\end{sidewaysfigure}


\begin{figure}
    \centering
    \includegraphics[width=\linewidth]{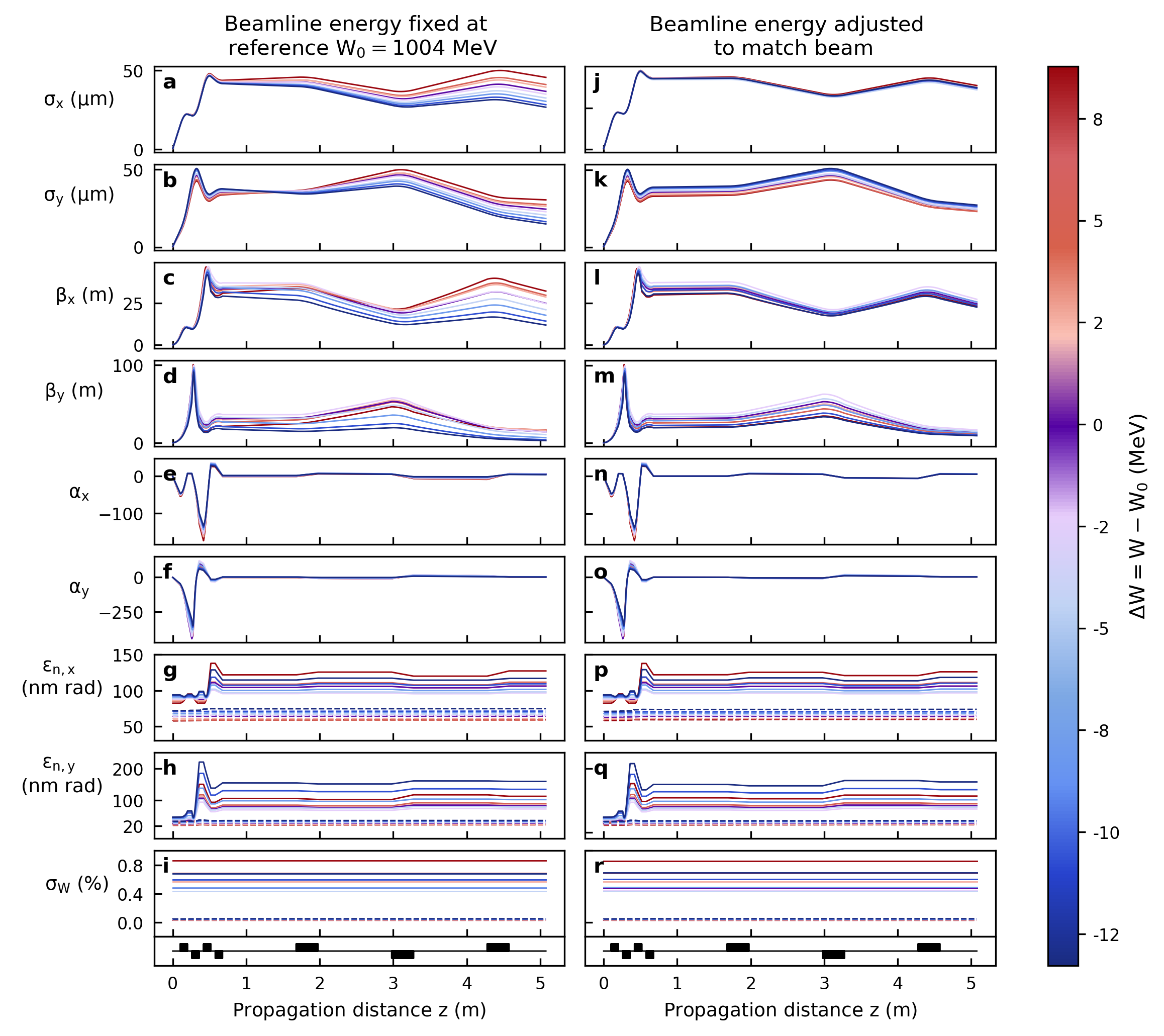}
    \caption*{\textbf{Extended Data Fig. 2} Witness beam parameters during propagation through the transport line for two cases. First column: beamline reference energy fixed at the energy of the reference witness beam with optimised beam-loading ($W_0 = 1004$\,MeV). Second column: beamline energy adjusted to match the energy of each individual witness beam (energies in Fig. \ref{fig2}a). Beams are colour-coded on mean energy with the reference beam in purple. \textbf{a, j,} rms radius in $x$ ($\sigma_x$). \textbf{b, k,} rms radius in $y$ ($\sigma_
    y$). \textbf{c, l,} Twiss beta in $x$ ($\beta_x$). \textbf{d, m,} Twiss beta in $y$ ($\beta_y$). \textbf{e, n,} Twiss alpha in $x$ ($\alpha_x$). \textbf{f, o,} Twiss alpha in $y$ ($\alpha_y$). \textbf{g, p,} Normalised projected emittance ($\varepsilon_\mathrm{n,p,x}$) (solid lines) and slice emittance ($\varepsilon_\mathrm{n,s,x}$) (dashed lines) in $x$.  \textbf{h, q,} Normalised projected emittance ($\varepsilon_\mathrm{n,p,y}$) (solid lines) and slice emittance ($\varepsilon_\mathrm{n,s,y}$) (dashed lines) in $y$.  \textbf{i, r,} Projected energy spread ($\sigma_\mathrm{W,p}$) (solid lines) and slice energy spread ($\sigma_\mathrm{W,p}$) (dashed lines).}
    \label{fig:ext_fig2}
\end{figure}

\clearpage

\begin{figure}
     \centering
    \includegraphics[height=0.75\textheight]{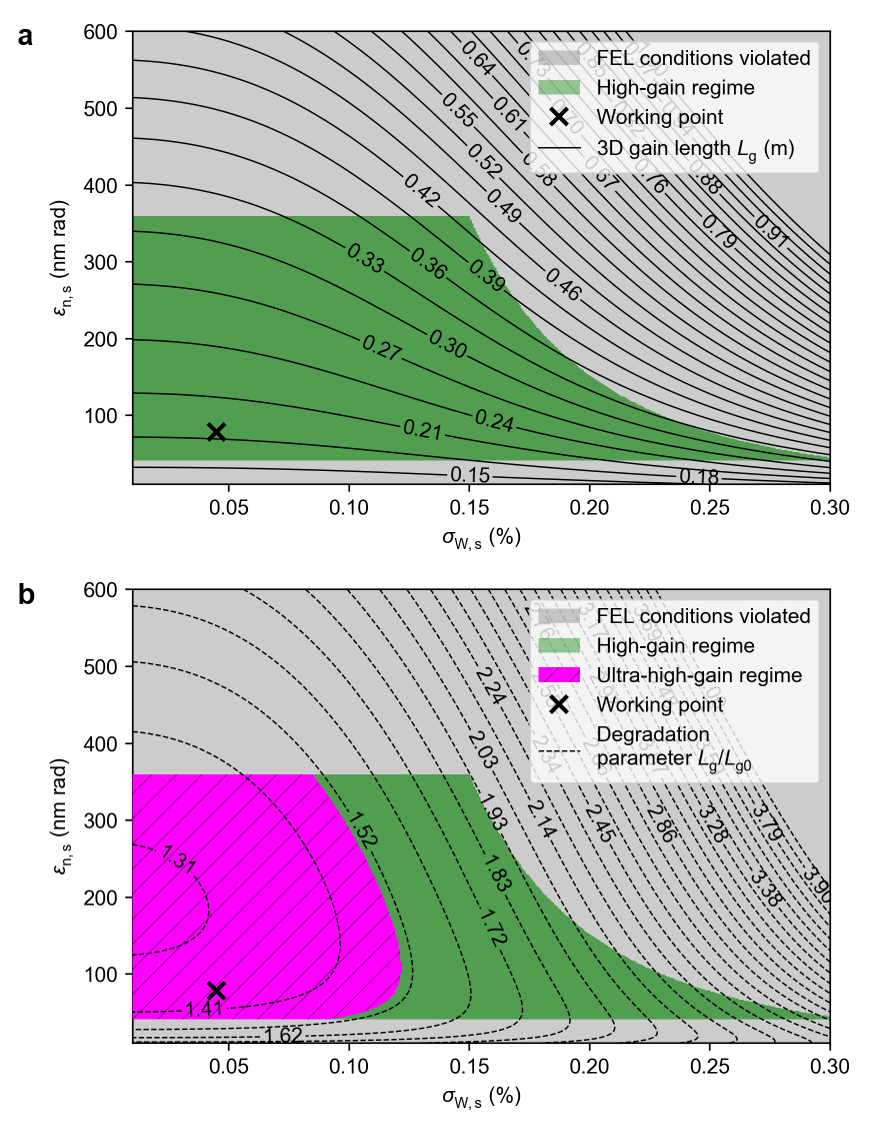}

    \caption*{\textbf{Extended Data Fig. 3:} The Xie parametrisation can be used to estimate the FEL power gain length based on beam and undulator parameters \cite{ming_xie_design_1995}. \textbf{a,} Contours show the estimated 3D power gain length $L_\mathrm{g}$ as a function of slice normalised emittance $\varepsilon_\mathrm{n,s}$ and relative slice energy spread $\sigma_\mathrm{W,s}$. The parameters of the reference beam ($I_\mathrm{pk}=5.3$\,kA, $W=1004$\,MeV) are used with $\beta^*=1$\,m. The black cross marks the working point ($\varepsilon_\mathrm{n,s}=78$\,nm\,rad and  $\sigma_\mathrm{W,s}=0.045$\,\%). Grey shading shows where the FEL lasing criteria would be violated, green shading shows where they are fulfilled. The contours show that the working point is far from a cliff edge and that the gain length $L_\mathrm{g} \sim0.20$\,m. \textbf{b,} Contours show the degradation parameter $L_\mathrm{g}/L_\mathrm{g0}$, where $L_\mathrm{g0}$ is the ideal 1D power gain length. We use this parameter to define the `ultra-high-gain regime', where $L_\mathrm{g}/L_\mathrm{g0} \lessapprox 1.5$ as shown by the pink shading.}
        \label{ext_fig3}
\end{figure}

\clearpage

\begin{figure}
    \centering
    \includegraphics[width=\linewidth]{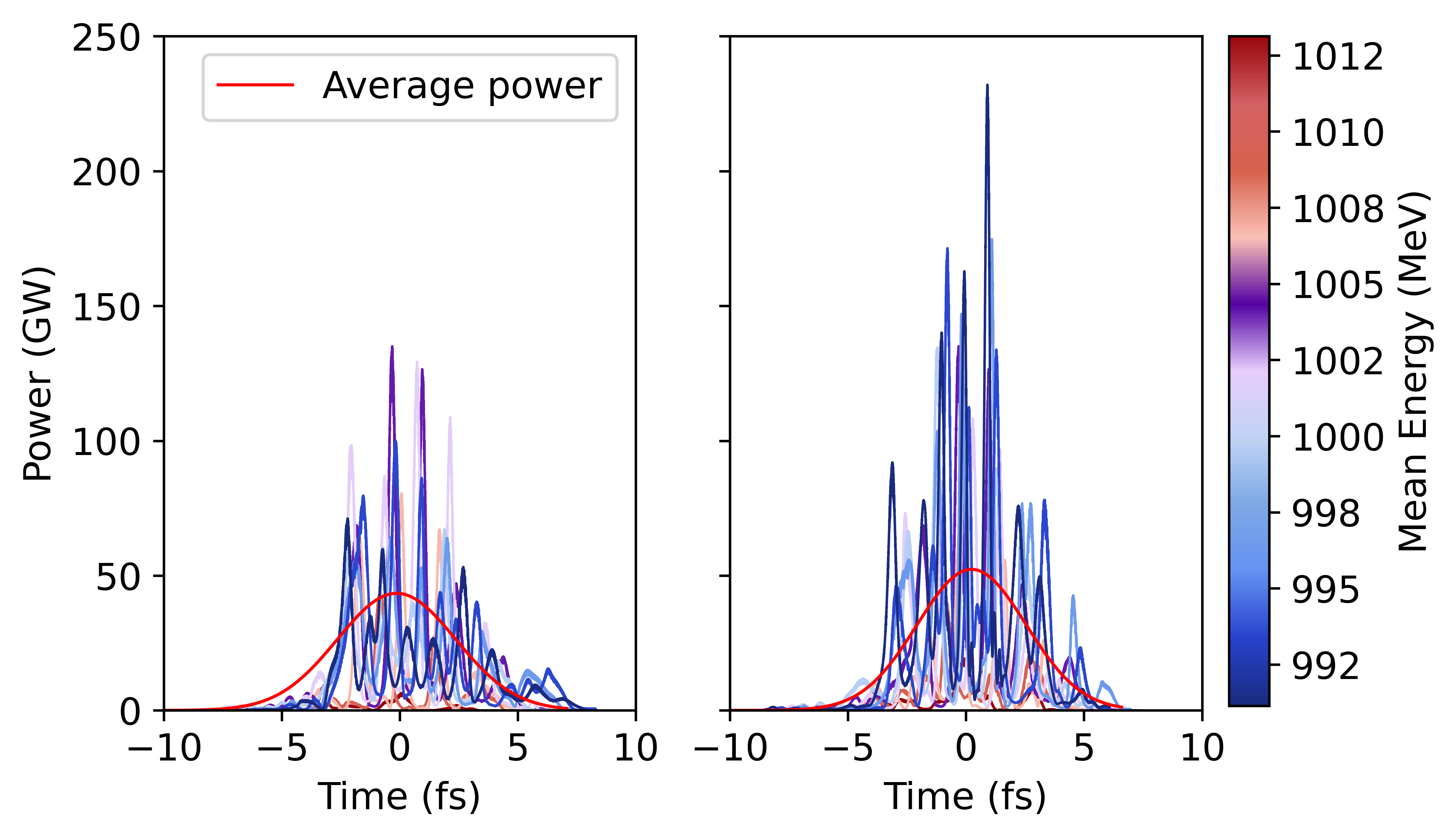}
    \caption*{\textbf{Extended Data Fig. 4:} Pulse profiles near saturation for the fixed energy beamline (left) and adjusted energy beamline (right). Pulses are colour-coded by witness beam energy. Red lines show a Gaussian fit to the average power. In the fixed beamline case the averaged peak power is 43\,GW with FWHM pulse duration of 5.9 fs. In the adjusted energy beamline case  the averaged peak power is 52\,GW with FWHM pulse duration of 5.4 fs. }
    \label{ext_fig4}
\end{figure}

\newpage

\end{document}